\RequirePackage{fix-cm}
\documentclass[smallextended]{svjour3}      
\smartqed  
\usepackage{graphicx}
\usepackage{amsmath}  
\usepackage{cite}
\usepackage{graphicx}
\usepackage{tcolorbox}
\usepackage{amsfonts}
\usepackage{tikz}
\usepackage{array}
\newcolumntype{M}[1]{>{\centering\arraybackslash}m{#1}}
\begin{document}

\title{Non-locality and entanglement in multi- qubit systems from a unified framework
}

\author{Sooryansh Asthana        \and
        Soumik Adhikary \and V. Ravishankar %etc.
}

\institute{Sooryansh Asthana  \at Department of Physics, Indian Institute of Technology Delhi, New Delhi-110016, India, \\\email{sooryansh.asthana@physics.iitd.ac.in} \and Soumik Adhikary \at Department of Physics, Indian Institute of Technology Delhi, New Delhi-110016, India, \\\email{soumikadhikary@physics.iitd.ac.in} \and  V. Ravishankar
 \at Department of Physics, Indian Institute of Technology Delhi, New Delhi-110016, India, \\ \email{vravi@physics.iitd.ac.in}}

\date{Received: date / Accepted: date}

\maketitle

\begin{abstract}
Non-classical probability is the underlying feature of quantum mechanics. The emergence of Bell-CHSH non-locality\footnote{We stress that CHSH inequality\cite{Clauser69} is more general and reduces to Bell inequality\cite{Bell64} under the assumption of perfect correlations between the observables.} for bipartite systems and linear entanglement inequalities for two-qubit systems has been shown in Adhikary {\it et al. 2020} [Eur. Phys. J. D 74, 68 (2020)], purely as  violations of classical probability rules.  In this paper, we improve upon that work by showing that violation of any nonlocality inequality implies violation of classical probability rules, manifested through negative probabilities, without recourse to any underlying theory. Moving on to entanglement, we employ parent pseudoprojections to show  how any number of linear and nonlinear entanglement witnesses for multiqubit systems can be obtained as violations of classical probability rules. They include the ones that have been derived earlier by employing different methods. It provides a perspective complementary to the current understanding in terms of the algebraic approaches.
\end{abstract}

\section{Introduction} 
\label{Introduction}
In quantum information, endeavour for identification of non-classical states has been motivated mainly by the non-classical applications that they offer. These applications are either altogether novel \cite{Bennett93, Ekert91} or show advantage over their classical counterparts \cite{Deutsch85}. Promising though these applications are, the true potential in various practical scenarios, such as conference quantum key distribution and multi-party quantum communication tasks can unravel in multi-party scenarios. So, various multi-partite generalisations of quantum communication protocols have been shown to this end  \cite{Bouda01, Jian07, Epping17}. Non-classicality of correlations, e.g.,  non-locality \cite{Svetlichney87}, quantum entanglement \cite{Horodecki09} and quantum discord \cite{Ollivier01} acts as a resource in these applications. It has led to many criteria for detection of multiparty states possessing different features of nonclassicality \cite{Svetlichney87, Mermin90, Acin01, Seevinck02,  Toth05, Guhne10, Vicente11}. The criteria are motivated from distinctive features of resource states (e.g., nonlocal and entangled states) vis-a-vis the so-called free states (e.g., local and separable states).  Since there are many apparently different paradigms of (non-) classicality, the need for a common framework from which all the criteria of non-classicality are emergent becomes paramount.

Needless to mention,  such a framework should rest on fundamental properties of quantum mechanics. As quantum mechanics is a new theory of probability \cite{Dirac42, Bartlett45, Feynman87}, violation of classical probability rules becomes an important avenue for emergence of various non-classicality criteria. In his seminal paper \cite{Fine82}, Fine has shown that the existence of joint probability distribution is equivalent to the condition of existence of hidden variable model, i.e., Bell inequality holds. Thus, nonlocality, a nonclassical feature not restricted only to quantum mechanics, can also be looked upon as the non-existence of joint probability distribution. Along the same lines, we show in this paper how nonlocality ineqaulities can be derived from violations of classical probability rules.
We show that nonlocality leads to pseudoprobability scheme with negative entries (probabilities). This does not take recourse to any underlying physical theory. In fact, we present a formal proof that violation of any nonlocality inequality can be looked upon as a violation of classical probability sum rule.
Existence of a nonnegative joint probability scheme would have precluded such a nonclassical behaviour.

 Since entanglement is a nonclassical feature of quantum mechanics, in order to lay down an operational framework incorporating nonclassical probability, we  introduced, in a  recent work \cite{Adhikary20}, a new class of operators called pseudo projections. They are, by definition, quantum representatives of indicator functions for classical events in phase space. We showed that their expectation values -- for  a  given state -- have the significance of  pseudo probabilities. When pseudo probabilities take values outside the interval $[0,1]$, they flag a non classical property of the underlying state. For, they essentially exhibit violations of classical probability rules. 

A special case of interest is the Margenau Hill distribution \cite{Hill61, Barut88},  which is just  the pseudo projection for the joint outcome of two observables (pseudo projections for outcomes of a single observable are always projections.). Employing just these minimal operators, we have recovered, in \cite{Adhikary20},  a number of non-classicality criteria which have been derived earlier from diverse considerations. They include Bell-CHSH inequality in any dimension and entanglement inequalities for two qubit states. In a subsequent work \cite{Asthana20}, we have further established the relationship between anomalous weak values and pseudo- probabilities. We have also derived conditions for quantum coherence in single qubit systems and quantum discord in two qubit systems. The generic nature of the framework gets manifested in that the criteria for so many different features of nonclassicality emerge from it as  violations of classical probability rules. 

In this paper, as a direct continuation of \cite{Adhikary20, Asthana20},  we answer the questions: (i) whether conditions for nonlocality and entanglement in multiparty  and multiqubit systems respectively can also be looked upon as violations of classical probability rules and, (ii) how violation of classical probability rules can yield nonlinear entanglement inequalities in multi-qubit systems?

We employ our approach to unravel the violation of classical probability rules when the multi-party nonlocality inequalities derived in \cite{Svetlichney87, Mermin90, Seevinck02, Das17} get violated. Next, we show that multi-qubit entanglement inequalities, obtained in \cite{Acin01, Toth05, Guhne10} through various approaches, also emerge as violation of classical probability sum rules by employing pseudoprojections. 
In summary, we show that nonlocality leads to negative probabilities without any recourse to any physical theory, whereas entanglement leads to negative pseudoprobabilities constructed according to quantum mechanical rules.

The plan of the paper is as follows: For clarity and completeness, a brief account of the formalism for nonlocality and entanglement will be given in sections (\ref{Preliminaries_Nonloc}) and (\ref{Preliminaries}) respectively.  We discuss relation of our work with the previous works in section (\ref{Relation_previous}). In section (\ref{Proof_NL}), we prove a central result, {\it viz.}, violation of any condition for locality is concomitant on violation of a classical probability rule.  We set up the notations in section (\ref{Notation}) and explicitly identify pseudoprobabilities that turn negative for various multiparty nonlocality inequalities in section (\ref{Non-loc}). Section (\ref{ent}) contains the results on entanglement. After describing  the methodology to choose pseudoprobabilities for deriving entanglement inequalities in section (\ref{Methodology}), we present entanglement inequalities in sections (\ref{EntLin3}- \ref{Ent_linN}). In section (\ref{Illustration}), relative strengths of various inequalities are compared. Finally, section (\ref{Conclusion}) summarises the paper.
  %---------------------------------------------

\section{The Formalism}
In this section, we lay down the formalism to be employed for deriving nonlocality and entanglement inequalities as violations of classical probability rules. We first explicitly state the methodology employed for nonlocality.
\subsection{Nonlocality}
\label{Preliminaries_Nonloc}
Nonlocality is a nonclassical feature not restricted to quantum mechanics. We start with a joint probability scheme without recourse to any physical model.  We add suitably chosen entries from this scheme. Classically, this sum would have been nonnegative.  We show that the conditions for nonlocality emerge when entries in the joint probability scheme turn negative.  We term the joint probabilities as pseudoprobabilities, as they can take negative values.
 In section (\ref{Proof_NL}), we prove that violation of any  nonlocality inequality always entails a sum of entries of underlying joint probability scheme to assume negative value. 
\subsection{Entanglement}
\label{Preliminaries}
In this section, we present the formalism to be employed for deriving sufficiency conditions for multiqubit entanglement. The pseudoprobabilities that we employ for obtaining entanglement inequalities are constructed using quantum mechanical rules.

The mathematical tool required for the formalism has been laid down  in \cite{Adhikary20}.   We recapitulate the formalism here for an uncluttered discussion. 
\subsubsection{Classical probability for an outcome of a single observable}
Let there be a classical system in state $f$ and  $M$ represents an observable  taking values from the set $\{m_i; i=1,2,\cdots\}$. In classical probability theory, in order to find the probability of an event $M=m_i$, one needs to identify the support $S^M_{m_i}$ for the event. Support is that region of event space in which the probability of happening of that event is 1. Thereafter, the overlap of the indicator function, $I_{S^M_{m_i}}$, constructed over the support  $S^M_{m_i}$, with the state $f$ yields the probability of the event. Indicator functions are Boolean observables that assume value 1 inside the support and 0 outside it.
\subsubsection{Extension to two observables}

This method extends to more than one observable as well. For example, let there be two observables $M, N$ taking values in the sets $\{m_i; i=1,2,\cdots\}$ and $\{n_j; j=1,2,\cdots\}$.  The overlap of the indicator function constructed over the intersection of the supports of the two events, ${S^M_{m_i}}$ and ${S^N_{n_j}}$, with the state $f$ yields the probability for the event $M=m_i$ and $N=n_j$.  

\subsubsection{Quantum representatives of indicator functions for joint events: Pseudoprojections}
While transiting to quantum mechanics, observables map to hermitian operators.  Indicator functions for different outcomes of a given observable map to  projections of the hermitian operators with the corresponding eigenvalues.  The crux of the matter is that indicator functions for joint outcomes of two or more observables do not map to projection operators, unless the observables commute.  Such indicator functions for the joint outcomes of multiple observables map to pseudo projections \cite{Adhikary20}, which are symmetrised products of the individual projections. By construction, pseudo projections are hermitian, but not idempotent. Nor are their spectrums bounded in the unit interval $[0,1]$, unlike their parent indicator functions.  Violation of this classical bound is the source of non-classicality  in quantum mechanics.

 Consider, for instance, the two observables $M, N$  and let $\pi_{m_i},~\pi_{n_j}$
be the projection operators representing the respective indicator functions  for the outcomes $M = m_i$ and $N = n_j$.  The operator, i.e., the {\it  pseudo-projection} representing the classical indicator function
 for their  joint outcome is the symmetrised product \cite{Hill61, Barut88}:
 \begin{equation}
 \mathbf{\Pi}_{m_i n_j} = \frac{1}{2}\Big\{ \pi_{m_i},\pi_{n_j} \Big\},
 \end{equation}
  in accordance with Weyl ordering \cite{Weyl27}.  %Generalization to multiple projections is not unique and is a source of much richness.
  \subsubsection{Pseudoprojection for joint outcomes of more than two observables}
  Pseudoprojections representing joint outcomes of more than two observables can also be constructed similarly. For example, consider the joint outcomes,  $O_1=o_1, \cdots, O_N=o_N$, of $N$ observables. If $\pi_{o_i}$ represents the projection for $O_i=o_i$, their product can be permuted, in general, in $N!$ ways. If all the projections happen to be distinct from each other, this will give rise to $N!/2$ quantum representatives of the form
  \begin{align}
     {\bf \Pi}_{o_1\cdots o_N} =\dfrac{1}{2}\big(\pi_{o_1}\cdots \pi_{o_N}+{\rm h.~c.}\big), 
  \end{align}
where ${\rm h.c.}$ represents hermitian conjugate. A pseudoprojection obtained from a given order of $N$ distinct projections is termed as {\it unit pseudoprojction}.

 All $N!/2$ unit pseudoprojections are legitimate quantum representatives of the same classical indicator function. For this reason, one may consider their convex sums, which will also be equally valid quantum representatives of classical indicator function\footnote{ This ambiguity is the same as we see, e.g., for the classical function $x^2p$, which can be represented as $\hat{x}^2\hat{p}, \hat{p}\hat{x}^2, \hat{x}\hat{p}\hat{x}$ or as their linear combinations in quantum mechanics.}.
\subsubsection{Pseudoprobability in quantum mechanics}

The overlap of a pseudoprojection with a state $\rho$ is defined to be pseudo-probability, i.e., the pseudo-probability ${\cal P}_{o_1\cdots o_N}$ for joint outcomes of $O_1=o_1, \cdots, O_N=o_N$ in a state $\rho$ is defined as, 
\begin{align}
 {\cal P}_{o_1\cdots o_N} \equiv {\rm Tr} ( {\bf \Pi}_{o_1\cdots o_N}\rho). 
\end{align}

    Pseudoprojections for a multipartite system are direct products of those for single subsystems.

 Only a  pseudo-projection involving mutually commuting projections is a true projection, whence   its expectation value in any state  necessarily possesses  an interpretation of being a probability.  More generally,  if we were to construct a scheme of joint pseudo-probabilities (PPS), involving  all possible outcomes of each observable, then only the marginals, involving only sets of mutually commuting  observables, are guaranteed to have the character of a  classical probability scheme -- with entries in agreement with the predictions of quantum mechanics. 

This leads to a broad definition of non-classicality of a state with respect to a given set of observables \cite{Adhikary20}.

\subsubsection{Nonclassicality of quantum states}
{\it Definition:} A quantum state is nonclassical if even {\it one} pseudo-probability in the PPS is negative.

 This definition is more general than the ones given in \cite{Hill61,  Barut88, Johansen_04, Pusey14} since it incorporates pseudo-probabilities generated by pseudo-projections for joint outcomes of any number of observables and by convex sums of unit pseudoprojections as well. The sum of a chosen set of pseudoprobabilities of a scheme may also assume value outside [0, 1] and may act as an independent signature of nonclassicality\footnote{Without getting into further intricacies, one may define pseudo-projection for operations such as OR and NOT by using standard Boolean rules which have been worked out in \cite{Adhikary20}.
}.

Indeed, PPS, is of  fundamental importance. If all the entries in a PPS turn  to be non-negative, it will be equivalent to a classical joint probability scheme; there would be no non classicality. This leads us to conjecture that pseudoprobabilities capture all the nonclassical features of quantum mechanics. If this be so, one must recover the conditions for various manifests of nonclassicality in quantum mechanics, in particular, entanglement in multi-qubit systems from pseudoprobabilities. In section (\ref{ent}), we show that it is indeed so.

 Before we move to our main results, we make a comparison between our work and the previous works in the next section.
\section{Relation with previous works}
\label{Relation_previous}
Historically, we note that  the product of noncommuting projections was first employed by Kirkwood \cite{Kirkwood33} and Barut \cite{Barut57} for construction of a complex probability distribution. Later, Margenau and Hill \cite{Hill61}, and Barut et al.\cite{Barut88} employed the hermitized  product  for construction of a quasiprobability distribution of joint outcome of non-commuting operators. Such a distribution has been used in \cite{Puri12} to characterise nonclassical correlations of multiqubit system with prior knowledge of average direction of single subsystem. 
The present approach starts with the more fundamental indicator functions, and  systematically provides conditions for non-classicality purely as violations of classical probability rules.  This, to the best of our knowledge, has not been explored before.

 In this paper, we are interested in two specific forms of nonclassicality: nonlocality in multiparty systems and entanglement in multiqubit systems. The conditions for multiparty nonlocality have earlier been derived assuming (i) hybrid local-nonlocal models \cite{Svetlichney87, Seevinck02} and (ii) completely factorisable local hidden variable model \cite{Mermin90}.  In this paper, we show that conditions for nonlocality emerge when there are negative entries in the underlying joint probability scheme. The two approaches are dual to each other, as shown by Fine for Bell inequality \cite{Fine82}. In fact, our approach gives a mathematical framework to the hidden variable models and those models provide a physical basis for our appraoch.  

Conditions for different kinds of multiqubit entanglement have earlier been obtained through (i) conditions on entries of density matrices of biseparable and completely separable states vis-a-vis entangled states \cite{Guhne10}, (ii) stabiliser formalism \cite{Toth05}, (iii) witnesses involving projection operators for entangled states \cite{Acin01}, to name a few. In contrast, in our work, we show that all these conditions can be arrived at purely by looking at violation of classical probability rules through pseudoprobabilties in quantum mechanics.
  %---------------------------------------------
 \section{Violation of any linear nonlocality inequality implies violation of a classical probability rule}
 \label{Proof_NL}
 In this section, we prove that violation of any linear nonlocality inequality is equivalent to the nonexistence of an underlying classical joint probability scheme. We show it for a bipartite system but the proof admits a straightforward generalisation to multipartite systems as well.
 \subsection{Condition for locality}
 Let $\{C_{\alpha}\}$ and $\{D_{\beta}\}$ be sets of $N_1$ and $N_2$ observables for the first and the second subsystems of a bipartite system respectively. The respective sets of outcomes of the observables $C_{\alpha}$ and $D_{\beta}$ are represented by $\{c_{\alpha}^{i_{\alpha}}\}$ and $\{d_{\beta}^{j_{\beta}}\}$. If ${\cal P}(C_{\alpha}=c^{i_{\alpha}}_{\alpha}; D_{\beta}=d_{\beta}^{j_{\beta}})$ represents the  probability for the joint event $C_{\alpha}=c_{\alpha}^{i_{\alpha}}$ and $D_{\beta}=d_{\beta}^{j_{\beta}}$, a linear nonlocality inequality obeyed by local hidden variable models \cite{Brunner14} would read as follows:
 \begin{align}
 \label{Nonlocality}
     {\cal I}\equiv \sum_{\alpha\beta}\sum_{i_{\alpha}j_{\beta}}w^{\alpha\beta}_{i_{\alpha}j_{\beta}}{\cal P}(C_{\alpha}=c^{i_{\alpha}}_{\alpha}; D_{\beta}=d^{j_{\beta}}_{\beta}) \leq  R.
 \end{align}
 We can take all the coefficients $ w^{\alpha\beta}_{i_{\alpha}j_{\beta}} \geq 0$, for if any of them were negative, we rewrite the corresponding ${\cal P}$ as $1-{\cal P}'$, where ${\cal P}'$ is the probability for the complementary event. This implies the existence of an upper bound $R \geq 0$ for all local hidden variable models.

\noindent{\it Lemma:} Let ${\cal M}_1, {\cal M}_2, \cdots$ be the sets of all mutually consistent events  $\{C_{\alpha}=c^{i_{\alpha}}_{\alpha};  D_{\beta}=d_{\beta}^{j_{\beta}}\}$, and let ${\cal N}_1, {\cal N}_2, \cdots$ represent the sets of corresponding coefficients $w^{\alpha\beta}_{i_{\alpha}j_{\beta}}$ in inequality $(\ref{Nonlocality})$. Mutually consistent events imply the events that can be simultaneously assigned probability equal to one in a local hidden variable model. If, for a given ${\cal N}_k$, the sum of all coefficients $w^{\alpha\beta}_{i_{\alpha}j_{\beta}} \in {\cal N}_k$ is $R_k$, then,
\begin{align}
\label{Lemma}
R \geq \underset{\{{\cal N}_k\}}\max R_k,
\end{align}
where  the maximum is taken over all ${\cal N}_1, {\cal N}_2, \cdots$.\\

\noindent{\it Proof:} For a given $k$, there always exists a local hidden variable model in which all the events belonging to the set ${\cal M}_k$ can be assigned unit probability, and all the events inconsistent with the events belonging to ${\cal M}_k$ have zero probability. Since $k$ is an arbitrary label and the inequality (\ref{Nonlocality}) is obeyed by all the local hidden variable models,  equation (\ref{Lemma}) holds.\\

 \noindent{\it Theorem 1: Violation of the inequality  (\ref{Nonlocality}) implies violation of a classical probability rule.}\\
\noindent{\it Proof:} We prove by contradiction. Assume that there exists an underlying nonnegative joint probability scheme,
\begin{align}
 \Big\{{\cal P}\Big(C_1=c_1^{i_1}, \cdots, C_{N_1}=c_{N_1}^{i_{N_1}};  D_1=d_1^{j_1}, \cdots, D_{N_2}=d_{N_2}^{j_{N_2}}\Big)\Big\},\nonumber
\end{align}
which we compactly represent as $\{{\cal P}\big(\{C_{\alpha}=c_{\alpha}^{i_{\alpha}}\};   \{D_{\beta}=d_{\beta}^{j_{\beta}}\}\big)\}$.
 Consider the expression
\begin{align}
    R-{\cal I} =&R-\sum_{\alpha\beta}\sum_{i_{\alpha}j_{\beta}}w^{\alpha\beta}_{i_{\alpha}j_{\beta}}{\cal P}(C_{\alpha}=c_{\alpha}^{i_{\alpha}}; D_{\beta}=d_{\beta}^{j_{\beta}}),\nonumber
    \end{align}
    which we rewrite as $R\cdot 1-{\cal I}$, where we insert the sum rule for joint probabilities,
    \begin{align} \sum_{\substack{i_{1}\cdots i_{N_1}\\j_1\cdots j_{N_2}}}{\cal P}\big(\{C_{\alpha}=c_{\alpha}^{i_{\alpha}}\};  \{D_{\beta}=d_{\beta}^{j_{\beta}}\}\big)=1,
    \end{align}
    Thus, 
    \begin{align}
&R-{\cal I} \nonumber\\   &=R\sum_{\substack{i_{1}\cdots i_{N_1}\\j_1\cdots j_{N_2}}}{\cal P}\big(\{C_{\alpha}=c_{\alpha}^{i_{\alpha}}\};  \{D_{\beta}=d_{\beta}^{j_{\beta}}\}\big)
    -\sum_{\alpha\beta}\sum_{\substack{i_1\cdots i_{N_1}\\j_1\cdots j_{N_2}}}w^{\alpha\beta}_{i_{\alpha}j_{\beta}}{\cal P}\big(\{C_{\alpha}=c_{\alpha}^{i_{\alpha}}\}; \{D_{\beta}=d_{\beta}^{j_{\beta}}\}\big).
  \label{Nonlocalityexpansion}
    \end{align}    
The proof follows by employing the inequality in equation (\ref{Lemma}). Since only mutually consistent events in a local hidden variable model have common joint events, the coefficient of a given joint probability ${\cal P}\big(\{C_{\alpha}=c_{\alpha}^{i_{\alpha}}\};  \{D_{\beta}=d_{\beta}^{j_{\beta}}\}\big)$ in the second term of equation (\ref{Nonlocalityexpansion}) equals $R_k$, for some $k$. So, equation (\ref{Nonlocalityexpansion}) contains the sum of joint probabilities with nonnegative coefficients. If, in  addition, all the joint probabilities were to be nonnegative, it would follow that,
\begin{align}
    R-\sum_{\alpha\beta }\sum_{i_{\alpha}j_{\beta} }w^{\alpha\beta}_{i_{\alpha}j_{\beta}}{\cal P}(C_{\alpha}=c_{\alpha}^{i_{\alpha}};D_{\beta}=d_{\beta}^{j_{\beta}}) \geq 0,
\end{align}
which agrees with the locality condition (\ref{Nonlocality}). Thus, a violation of inequality ($\ref{Nonlocality}$) is possible only if the weighted sum of joint probabilities
given in equation (\ref{Nonlocalityexpansion}) is negative. This, in turn, implies that in place of joint probabilities, we have  pseudoprobabilities that assume negative values as well. The validity of the result is not restricted to linear nonlocality inequalities. That it can be extended to nonlinear nonlocality inequalities is shown in Appendix (\ref{NonlinearNonlocProof}).

Theorem 1 and 1A (proved in Appendix (\ref{NonlinearNonlocProof})) provide the formal basis for our constructions of combinations of pseudoprobabilities to exhibit violations of classical probability rules. In this paper, we shall focus our attention exclusively on nonlocality and entanglement inequalities in multiparty and multiqubit systems respectively.
\section{Notations}
\label{Notation}
In this section, we setup  compact notations to be used henceforth in the paper for expressing the results. This is made possible because we  employ only dichotomic observables, with outcomes $\pm 1$.  
\begin{enumerate}
\item The symbol ${\cal E}$ is reserved to denote events.
\item Recall that the symbol ${\cal P}$ has a dual significance: it represents pseudoprobabilities for joint outcomes of observables  for nonlocality, which can assume both negative and nonnegative values. For entanglement, it represents  pseudo probabilities, defined as expectations of the pseudoprojections for the corresponding joint events.
\item    $\mathcal{E}(A=+1) \equiv \mathcal{E}(A)$, and its negation, $\mathcal{E}(A =-1) \equiv \mathcal{E}({\bar A})$. 
\item   For an  $N$ party system, the observable $A_i$ refers to the $i^{th}$ subsystem. Observables belonging to the same subsystem are distinguished by primes in the superscript, such as $A'_i,~A''_i,\cdots$. 
\item  The pseudo-probability for a joint event ${\cal E}(A_1 = +1, A'_1 = +1; A_2 = +1, A'_2 = -1)$ is represented compactly as $\mathcal{P}(A_1 A'_1; A_2 {\bar A}'_2)$. This rule admits  easy generalisation. 
\item The pseudo probability 
$\mathcal{P}(A_1=A'_1)$ represents the sum  $\mathcal{P}(A_1 A'_1) +  \mathcal{P}(\bar{A}_1 \bar{A}'_1)$.
\item  The  compact notation  ${\cal P} ({\cal E}; A)$ represents the pseudo-probability for the joint  occurrence of event ${\cal E}$ together with $A=+1$.
\item The symbol $\vee$ is employed for standard OR operation. Since, we use mutually exclusive events, so
\begin{align}
    {\cal P}({\cal E}_1\vee {\cal E}_2) ={\cal P}({\cal E}_1)+{\cal P}({\cal E}_2),\nonumber
\end{align}
for any two events ${\cal E}_1$ and ${\cal E}_2$.
\item As an exception, we employ  lower case letters for the observables of a qubit:  $A_i \rightarrow a_i \equiv \vec{\sigma}_i\cdot\hat{a}_i$. 
\end{enumerate}
  %---------------------------------------------
  
  %---------------------------------------------

\section{Multiparty non-locality}
\label{Non-loc}
 Equipped with the result of section (\ref{Proof_NL}), we apply the framework to multiparty nonlocality.  Combinations of appropriate pseudoprobabilities of a joint pseudoprobability scheme yield us the required inequality. Furthermore, we reexpress the pseudoprobabilities in terms of  expectations of observables, that can be experimentally measured.
\subsection{Svetlichny Inequality}
Svetlichny inequality detects $N$-- party nonlocal correlations that are  not reducible to ($N-M$) party correlations related locally to the rest of $M$--parties ($M<N$).
For illustration, we consider  the simplest  case of three party system first. 
\subsubsection{N=3}
 Of relevance to this example is the specific sum of pseudoprobabilities, given by
\begin{align}
\label{Svetlichny}
{\cal P}_3^S \equiv &\mathcal{P}({\cal E}_2^{\alpha};A'_3)+ \mathcal{P}({\cal E}_2^{\alpha'}; {\bar A}'_3)
+ \mathcal{P}({\cal E}_2^{\beta};A_3)+\mathcal{P}({\cal E}_2^{\beta'} ; {\bar A}_3),
\end{align}
where the event ${\cal E}_2^{\alpha}$ is the  event underlying the Bell CHSH inequality for the residual two  party subsystem:
\begin{align}
\label{Bell}
{\cal E}_2^{\alpha} \equiv {\cal E}({\bar A}_1=A'_1=A_2)\vee{\cal E}(A_1=A'_1=A'_2).
\end{align}      
The  other three events, again for  the residual two party systems,  are obtained via the transformations of observables as shown in   table (\ref{Svet_3}).
 \begin{table}[ht]
 \centering
\begin{tabular}{|M{2.5cm}|M{4cm}|} 
 \hline
 Events &  Observables  \\ 
 \hline
 ${\cal E}_2^{\alpha} \rightarrow {\cal E}_2^{\beta}$  & Interchange of primed and unprimed observables\\ 
 \hline
 {${\cal E}_2^{\alpha} \rightarrow{\cal E}_2^{\alpha'}$}
  ${\cal E}_2^{\beta} \rightarrow{\cal E}_2^{\beta'}$  & {$A_2\leftrightarrow {\bar A}_2; ~A'_2\leftrightarrow {\bar A}'_2 $}\\
  \hline
\end{tabular}
  \caption{Transformation of events and corresponding transformation of observables.\label{Svet_3}}
 \end{table}
 
  We write the pseudoprobabilities in terms of expectations of observables by employing the sum rule for pseudoprobabilities and dichotomic nature of the observables (see Appendix (\ref{Svet_Calc}) for derivation).  The demand that ${\cal P}_3^S < 0$, which is forbidden classically, gives rise to the  inequality
\begin{align}
\label{Svetlichny1}
\langle S_3\rangle = &\Big\langle 
\big\{(-A_1+A'_1)A_2+ (A_1+A'_1)A'_2\big\}A'_3\nonumber\\
+&\big\{(A_1+A'_1)A_2+(A_1-A'_1)A'_2 \big\}A_3\Big\rangle < -4
\end{align}
If we now interchange ${\bar A}_1$ and ${\bar A}'_1$ with $A_1$ and $A'_1$ respectively in equation (\ref{Svetlichny}), we obtain the complementary condition
 $\langle S_3\rangle   > 4$
which, combined with
 (\ref{Svetlichny1}), yields the  simplest Svetlichny inequality,
\begin{align}
{\cal I}_{S_3}: ~\vert \langle S_3\rangle  \vert=&  \Big|\Big\langle \big\{(-A_1+A'_1)A_2+ (A_1+A'_1)A'_2\big\}A'_3\nonumber\\
+&\big\{(A_1+A'_1)A_2+(A_1-A'_1)A'_2 \big\}A_3\Big\rangle \Big| > 4
\end{align}

  %---------------------------------------------

\subsubsection{$N$-party system}
The above approach suggests  construction of pseudo probabilities  to set up    Svetlichny inequality for an $N$-- party system. It is identified to be the sum, recursively defined in terms of $(N-1)$ party events, given by
      \begin{align}
      \label{Svetlichny_N}
   {\cal P}^S_N \equiv &\mathcal{P}({\mathcal{E}_{N-1}^{\alpha}}; A'_N) + \mathcal{P} (\mathcal{E}_{N-1}^{\alpha'}; {\bar A}'_N)
    + \mathcal{P}(\mathcal{E}_{N-1}^{\beta}; A_N) + \mathcal{P}(\mathcal{E}_{N-1}^{\beta'}; {\bar A}_N); N \geq 3.
         \end{align}
      The associated event   ${\cal E}^{\alpha}_N$ is  given, recursively by
         \begin{align}
         {\cal E}^{\alpha}_N \equiv &  {\cal E}({\mathcal{E}_{N-1}^{\alpha}}; A'_N) \vee {\cal E}(\mathcal{E}_{N-1}^{\alpha'}; {\bar A}'_N)
          \vee {\cal E}(\mathcal{E}_{N-1}^{\beta}; A_N) \vee{\cal E}(\mathcal{E}_{N-1}^{\beta'}; {\bar A}_N); 
         \end{align}
         The  elementary event ${\cal E}_2^{\alpha}$   is defined in equation (\ref{Bell}). 
       The other events -- ${\cal E}_{N-1}^{\alpha'}, {\cal E}_{N-1}^{\beta}, {\cal E}_{N-1}^{\beta'}$,  can be obtained by changing the observables,  as prescribed in table  (\ref{Svet_N}).
         \begin{table}[ht]
  \begin{center}
\begin{tabular}{ |M{2.5cm}|M{4cm}| } 
 \hline
 Events &  Observables \\ 
 \hline
 ${\cal E}_{N-1}^{\alpha} \rightarrow{\cal E}_{N-1}^{\beta}$ & Change primed and unprimed observables.\\ 
 \hline
 ${\cal E}_{N-1}^{\alpha} \rightarrow{\cal E}_{N-1}^{\alpha'}$   ${\cal E}_{N-1}^{\beta} \rightarrow{\cal E}_{N-1}^{\beta'}$ & $A_{N-1}\leftrightarrow {\bar A}_{N-1}$ and $A'_{N-1}\leftrightarrow {\bar A}'_{N-1}$\\ 
  \hline
\end{tabular}
\end{center}
  \caption{Transformation of events and corresponding transformation of observables.\label{Svet_N}}
 \end{table}  
 
 After writing pseudoprobabilities in terms of expectations of observables,  the non classicality condition,  ${\cal P}^S_N<0$,  straightaway  yields the inequality (obtained in \cite{Bancal11}, by an entirely different method): 
 \begin{align}
{\cal I}_{S_N}: ~ \langle S_{N-1}A'_N+ S'_{N-1}A_N\rangle < -2^{N-1},
 \end{align}
 where $S_{N-1}$ is the  Svetlichny polynomial for the $(N-1)$--party  system;  $S'_{N-1}$ can be obtained from $S_{N-1}$ by interchanging the  primed and the unprimed observables.
  %---------------------------------------------

\subsection{Das-Datta-Agrawal inequalities.}
Recently, Das et al. \cite{Das17} derived an inequality for three party systems, by combining Bell inequalities for the constituent  two party subsystems. The inequality was tested against the so called generalised GHZ states, and subsequently generalised to multiparty states.  Here,  the stringent condition on correlations, imposed while deriving Svetlichny inequality, is relaxed to admit the existence of correlations involving $(N-1)$ parties as well.  As promised, we recover them as violations of appropriate classical probability rules. We start with the simplest case.
%------------------------------------------------

\subsubsection{$N$=3}
We start with  the   sum of pseudo-probabilities,
\begin{align}
{\cal P}^D_3 \equiv & {\cal P}({\cal E}_3^{\alpha})+{\cal P}({\cal E}_3^{\beta}),
\end{align}
for the events 
\begin{align}
\label{E1}
{\cal E}_3^{\alpha} & \equiv {\cal E}(A_1=A'_1=A_2; A_3)\vee{\cal E}(A_1 = A'_1={\bar A}_2; {\bar A}_3);\nonumber\\
{\cal E}_3^{\beta} & \equiv {\cal E}(A_1 = {\bar A}'_1=A'_2).
\end{align}
  Again after writing the pseudoprobabilities in terms of expectations of the observables,   the non-classicality condition,  ${\cal P}_3^D < 0$,  yields the inequality
\begin{align}
\label{Das_3}
& \big\langle (A_1 + A'_1)A_2A_3 +(A_1 - A'_1)A'_2\big\rangle < -2.
\end{align}
which, as we may see,  involves two party correlators also. Furthermore, it does not treat all the subsystems on par.
As in the previous subsection, the generalization to an $N$ party system is straightforward, except that we need to distinguish the two cases, {\it viz.},  $N$  even or odd.
%---------------------------------------------------------

\subsubsection{$N$ even}
The relevant sum of  pseudo probabilities is defined recursively,  by
\begin{align}
\label{Das_Neven}
 {\cal P}^D_N \equiv& {\cal P}({\cal E}_N^{\alpha})+{\cal P}({\cal E}_N^{\beta}); N\geq 4,
\end{align}
where 
\begin{align}
{\cal E}_N^{\alpha} & \equiv {\cal E}({\cal E}_{N-1}^{\alpha}; A_N)\vee {\cal E}({\cal E}_{N-1}^{\alpha'}; {\bar A}_N);\nonumber\\
{\cal E}_N^{\beta} & \equiv {\cal E}({\cal E}_{N-1}^{\gamma}; A'_N)\vee {\cal E}({\cal E}_{N-1}^{\gamma'}; {\bar A}'_N),
\end{align}
where  ${\cal E}_{N-1}^{\gamma} \equiv {\cal E}({\cal E}^{\beta}_{N-1}; A'_{N-1})\vee({\cal E}^{\beta'}_{N-1}; {\bar A}'_{N-1})$. The transformations  required to obtain ${\cal E}^{\alpha'}_{N-1}$, ${\cal E}^{\beta'}_{N-1}$ and ${\cal E}^{\gamma'}_{N-1}$ from ${\cal E}^{\alpha}_{N-1}, {\cal E}^{\beta}_{N-1}$ and ${\cal E}^{\gamma}_{N-1}$ respectively are shown  in table (\ref{Das_N}).
         \begin{table}[ht]
  \begin{center}
\begin{tabular}{ |c|c|c| } 
 \hline
 Events &  Observables \\ 
 \hline
 ${\cal E}_{N-1}^{\alpha} \rightarrow{\cal E}_{N-1}^{\alpha'}$ &$A_{N-1} \leftrightarrow {\bar A}_{N-1}$\\ 
 \hline
 ${\cal E}_{N-1}^{\beta} \rightarrow{\cal E}_{N-1}^{\beta'}$ & {$A'_{N-2} \leftrightarrow  {\bar A}'_{N-2}$}\\ 
 \hline
  ${\cal E}_{N-1}^{\gamma} \rightarrow{\cal E}_{N-1}^{\gamma'}$ & $A'_{N-1}\leftrightarrow {\bar A}'_{N-1}$\\ 
  \hline
\end{tabular}
\end{center}
  \caption{Transformation of events and corresponding transformation of observables.\label{Das_N}}
 \end{table} 
  
After plugging in the  expressions of pseudoprobabilities in terms of expectations, the condition ${\cal P}^D_N <0$, translates to the inequality
\begin{align}
\big\langle (A_1 + A'_1) A_2 \cdots A_N+ (A_1-A'_1)A'_2\cdots A'_N\big\rangle < -2.
\end{align}
It may be noted that every term has the full $N$ party correlation.

  %---------------------------------------------
\subsubsection{$N$ odd}
The pseudo probability  that we need is
\begin{align}
{\cal P}^D_N \equiv & {\cal P}({\cal E}^{\alpha}_{N})+{\cal P}({\cal E}^{\beta}_{N}); N \geq 5,
\end{align}
where the events,
\begin{align}
 {\cal E}^{\alpha}_N \equiv {\cal E}({\cal E}^{\alpha}_{N-1}; A_N)\vee {\cal E}({\cal E}^{\alpha'}_{N-1}; {\bar A}_N);
{\cal E}^{\beta}_N \equiv {\cal E}^{\beta}_{N-1}.
\end{align}
 The event ${\cal E}^{\alpha'}_{N-1}$ can be obtained from ${\cal E}^{\alpha}_{N-1}$ by the interchange $A_{N-1} \leftrightarrow {\bar A}_{N-1}$.
   After  writing the pseudoprobabilities in terms of expectations of observables and imposing the nonclassicality condition ${\cal P}^D_N$,  the ensuing  inequality  reads as
\begin{align}
\label{Das_Nodd}
\big\langle (A_1+A'_1)A_2 \cdots A_N+  (A_1-A'_1)A'_2\cdots A'_{N-1}\big\rangle < -2
\end{align}
which also gets a contribution from $(N-1)$ party correlators.

  %---------------------------------------------
\subsection{Mermin-Ardehali Inequality}
\label{Merm-Ard}
Mermin derived an inequality for a completely factorisable local hidden variable model for correlations of $N$--party states\cite{Mermin90}.  
We present the results, first for three party system and subsequently,
for $N$ party systems. These inequalities also differ for  $N$ even and $N$ odd, and   are expressed  in terms of  the  Mermin polynomial $M_N$, defined recursively by 
 $M_N = M_{N-1}A'_N +  M'_{N-1}A_N$,  with 
  $M_2 \equiv A_1A_2+A'_1A'_2$ and $M'_2 \equiv A_1A'_2-A'_1A_2$. $M'_N$ can be obtained from $M_N$ through the substitution: $A_{N}\rightarrow A'_N$ and $A'_N \rightarrow -A_N$. 

The inequality reads as
\begin{align}
\label{ma1}
{\cal I}_{M_N}: \langle M_N\rangle =\langle M_{N-1}A'_N+M'_{N-1}A_N\rangle < C_N
 \end{align}
where, $C_N = -2^{\frac{N}{2}}$   for $N$ even and $-2^{\frac{N-1}{2}}$ for $N$ odd. 
  %---------------------------------------------

\subsubsection{$N=3$}
  Following sum of pseudo-probabilities is of interest for us:
 \begin{align}
 \label{Mermin_3}
{\cal P}^M_3 \equiv &\mathcal{P}(\mathcal{E}_2^{\alpha}; A'_3) + \mathcal{P}(\mathcal{E}_2^{\alpha'}; {\bar A}'_3) 
+\mathcal{P}(\mathcal{E}_2^{\beta};A_3)+
\mathcal{P}(\mathcal{E}_2^{\beta'}; {\bar A}_3).
 \end{align}
Here, $\mathcal{E}_2^{\alpha} \equiv \mathcal{E} (A_1=A'_1=A_2=A'_2) \vee \mathcal{E} (A_1={\bar A}'_1=A_2={\bar A}'_2)$.  The transformations of observables required to obtain ${\cal E}_2^{\alpha'}, {\cal E}_2^{\beta}$ and ${\cal E}_2^{\beta'}$ are given in table (\ref{Merm_3}).
 \begin{table}[ht]
  \begin{center}
\begin{tabular}{ |M{2.5cm}|M{4cm}| } 
 \hline
 Events &  Observables  \\ 
 \hline
 ${\cal E}_2^{\alpha} \rightarrow{\cal E}_2^{\alpha'}$   ${\cal E}_2^{\beta} \rightarrow{\cal E}_2^{\beta'}$ & $A_2\leftrightarrow {\bar A}_2; ~A'_2\leftrightarrow {\bar A}'_2 $\\ 
 \hline
 ${\cal E}_2^{\alpha} \rightarrow {\cal E}_2^{\beta}$  & $A_2\rightarrow  A'_2; ~A'_2\rightarrow {\bar A}_2$ ${\bar A}_2\rightarrow {\bar A}'_2; ~{\bar A}'_2\rightarrow A_2$\\
  \hline
\end{tabular}
\end{center}
  \caption{Transformation of events and corresponding transformation of observables.\label{Merm_3}}
 \end{table}

 Writing the pseudoprobabilties in terms of expectations of observables and imposing the nonclassicality condition ${\cal P}_3^M <0,$ we arrive at equation (\ref{ma1}) with $N=3$.
%---------------------------------------------------------------------

\subsubsection{$N$ Even}
 We start with the sum
\begin{align}
{\cal P}^M_N & \equiv \mathcal{P}(\mathcal{E}_{N-1}^{\alpha}; A'_N )+\mathcal{P} (\mathcal{E}_{N-1}^{\alpha'}; {\bar A}'_N)
+\mathcal{P}(\mathcal{E}_{N-1}^{\beta}; A_N)+\mathcal{P}(\mathcal{E}_{N-1}^{\beta'}; {\bar A}_N); N \geq 4.
\end{align}
where the classical event${\cal E}^{\alpha}_N$ is recursively defined to be
\begin{align}
{\cal E}^{\alpha}_N \equiv & {\cal E}(\mathcal{E}_{N-1}^{\alpha}; A'_N) \vee {\cal E}(\mathcal{E}_{N-1}^{\alpha'}; {\bar A}'_N) \vee{\cal E}(\mathcal{E}_{N-1}^{\beta}; A_N)\vee {\cal E}(\mathcal{E}_{N-1}^{\beta'}; {\bar A}_N)
\end{align}
         \begin{table}[ht]
  \begin{center}
\begin{tabular}{ |M{2.5cm}|M{6cm}| } 
 \hline
 Events &  Observables  \\ 
 \hline
 ${\cal E}_{N-1}^{\alpha} \rightarrow{\cal E}_{N-1}^{\beta}$ & $A_{N-1} \rightarrow A'_{N-1}$ and $A'_{N-1} \rightarrow {\bar A}_{N-1}$;
 ${\bar A}_{N-1} \rightarrow{\bar A}'_{N-1}$ and ${\bar A}'_{N-1} \rightarrow A_{N-1}$\\
 \hline
 ${\cal E}_{N-1}^{\alpha} \rightarrow{\cal E}_{N-1}^{\alpha'}$
  ${\cal E}_{N-1}^{\beta} \rightarrow{\cal E}_{N-1}^{\beta'}$ & $A_{N-1}\leftrightarrow {\bar A}_{N-1}$ and $A'_{N-1}\leftrightarrow {\bar A}'_{N-1}$\\ 
  \hline
\end{tabular}
\end{center}
  \caption{Transformation of events and corresponding transformation of observables.\label{Merm_N}}
 \end{table}  
 The other events,  ${\cal E}_{N-1}^{\alpha'}, {\cal E}_{N-1}^{\beta}$ and ${\cal E}_{N-1}^{\beta'}$ are obtained via the substitutions and exchanges which are shown in table (\ref{Merm_N}).   Imposing the condition ${\cal P}_N^M <0,$ we arrive at equation (\ref{ma1}) with $C_N = -2^{N/2}$.
  %---------------------------------------------

\subsubsection{$N$ Odd}
 The sum of  pseudoprobabilities, which we require,  may be taken to be,
\begin{align}
 {\cal P}_N^M \equiv & {\cal P}({\cal E}_{N-3}^{\alpha'_1};{\cal E}_2^{\beta'_1};{ A}'_N)+{\cal P}({\cal E}_{N-3}^{\alpha'_2};{\cal E}_2^{\beta'_2};{ A}'_N)+
  {\cal P}({\cal E}_{N-3}^{\alpha_1};{\cal E}_2^{\beta_1};{ A}'_N)+{\cal P}({\cal E}_{N-3}^{\alpha_2};{\cal E}_2^{\beta_2};{ A}'_N)\nonumber\\
+&  {\cal P}({\cal E}_{N-3}^{\alpha'_1};{\cal E}_2^{\beta'_2};{\bar A}'_N)+{\cal P}({\cal E}_{N-3}^{\alpha'_2};{\cal E}_2^{\beta'_1};{\bar A}'_N)+
{\cal P}({\cal E}_{N-3}^{\alpha_1};{\cal E}_2^{\beta_2};{\bar A}'_N)+{\cal P}({\cal E}_{N-3}^{\alpha_2};{\cal E}_2^{\beta_1};{\bar A}'_N)\nonumber\\
+& {\cal P}({\cal E}_{N-3}^{\alpha'_1};{\cal E}_2^{\beta_1};A_N)+{\cal P}({\cal E}_{N-3}^{\alpha'_2};{\cal E}_2^{\beta_2};A_N)+
{\cal P}({\cal E}_{N-3}^{\alpha_1};{\cal E}_2^{\beta'_2};{\bar A}_N)+{\cal P}({\cal E}_{N-3}^{\alpha_2};{\cal E}_2^{\beta'_1};{ A}_N)\nonumber\\
+& {\cal P}({\cal E}_{N-3}^{\alpha'_1};{\cal E}_2^{\beta_2};{\bar A}_N)+{\cal P}({\cal E}_{N-3}^{\alpha'_2};{\cal E}_2^{\beta_1};{\bar A}_N)+
{\cal P}({\cal E}_{N-3}^{\alpha_1};{\cal E}_2^{\beta'_1};{\bar A}_N)+{\cal P}({\cal E}_{N-3}^{\alpha_2};{\cal E}_2^{\beta'_2};{\bar A}_N); N\geq 5.
\end{align}
where the events 
\begin{align}
&{\cal E}_{N-3}^{\alpha_1}\equiv{\cal E}({\cal E}_{N-5}^{\alpha'_1};{\cal E}_2^{\lambda'_1})\vee{\cal E}({\cal E}_{N-5}^{\alpha'_2};{\cal E}_2^{\lambda'_2})\vee
{\cal E}({\cal E}_{N-5}^{\alpha_1};{\cal E}_2^{\lambda_1})\vee{\cal E}({\cal E}_{N-5}^{\alpha_2};{\cal E}_2^{\lambda_2});\nonumber\\
&{\cal E}^{\alpha_1}_2\equiv {\cal E}(A_{1}=A'_{1}=A_{2}=A'_{2})\vee{\cal E}(A_{1}={\bar A}'_{1}=A_{2}={\bar A}'_{2});\nonumber\\
&{\cal E}_2^{\beta_1}\equiv {\cal E}(A_{N-2}=A'_{N-2}={\bar A}_{N-1}=A'_{N-1})\vee
{\cal E}({A}_{N-2}={ \bar A}'_{N-2}={\bar A}_{N-1}={\bar A}'_{N-1});\nonumber\\
&{\cal E}_2^{\lambda_1}\equiv {\cal E}({\bar A}_{N-4}=A'_{N-4}={ A}_{N-3}=A'_{N-3})\vee{\cal E}({\bar A}_{N-4}={\bar A}'_{N-4}=A_{N-3}={\bar A}'_{N-3}).
\end{align}
The transformations required to get the other events are given in table (\ref{Merm_Nodd}).
         \begin{table}[ht]
  \begin{center}
\begin{tabular}{ |M{2.5cm}|M{6cm}| } 
 \hline
Events &  Observables \\ 
 \hline
 ${\cal E}^{\alpha_1}_{N-3}\rightarrow{\cal E}^{\alpha'_1}_{N-3}$ & $A_{N-3}\rightarrow A'_{N-3}$; ${\bar A}_{N-3}\rightarrow {\bar A}'_{N-3}$; $A'_{N-3}\rightarrow {\bar A}_{N-3}$;  ${\bar A}'_{N-3}\rightarrow A_{N-3}$\\
 \hline
 ${\cal E}^{\alpha_1}_{N-3}\rightarrow{\cal E}^{\alpha_2}_{N-3};$ 
 ${\cal E}^{\alpha'_1}_{N-3}\rightarrow{\cal E}^{\alpha'_2}_{N-3}$ &$A_{N-3}\leftrightarrow {\bar A}_{N-3};$ $A'_{N-3}\leftrightarrow {\bar A}'_{N-3}$\\
 \hline
 ${\cal E}^{\beta_1}_{2}\rightarrow{\cal E}^{\beta_2}_{2}$ 
 ${\cal E}^{\beta'_1}_{2}\rightarrow{\cal E}^{\beta'_2}_{2}$ & $A_{N-2}\leftrightarrow {\bar A}_{N-2};$ $A'_{N-2}\leftrightarrow {\bar A}'_{N-2}$ \\
 \hline
 ${\cal E}^{\beta_1}_{2}\rightarrow{\cal E}^{\beta'_1}_{2}$ & $A_{N-1}\rightarrow {\bar A}'_{N-1};~{\bar A}_{N-1}\rightarrow {A}'_{N-1}$ ${A}'_{N-1}\rightarrow {A}_{N-1};~{\bar A}'_{N-1}\rightarrow {\bar A}_{N-1}$\\
 \hline
 ${\cal E}^{\lambda_1}_{2}\rightarrow{\cal E}^{\lambda'_1}_{2}$ & $A_{N-3}\rightarrow {\bar A}'_{N-3};~A'_{N-3}\rightarrow {A}_{N-3}$ ${\bar A}_{N-3}\rightarrow { A}'_{N-3};~{\bar A}'_{N-3}\rightarrow {\bar A}'_{N-3}$\\
 \hline
 ${\cal E}^{\lambda_1}_{2}\rightarrow{\cal E}^{\lambda_2}_{2}$  ${\cal E}^{\lambda'_1}_{2}\rightarrow{\cal E}^{\lambda'_2}_{2}$& $A_{N-4}\leftrightarrow {\bar A}_{N-4};~A'_{N-4}\leftrightarrow {\bar A}'_{N-4}$ \\

 \hline
\end{tabular}
\end{center}
  \caption{Transformation of events and corresponding transformation of observables.}\label{Merm_Nodd}
 \end{table} 
 
 As before, the non classicality condition 
 ${\cal P}^M_N< 0$, yields equation (\ref{ma1}), but with $C_N= -2^{\frac{N-1}{2}}$.
This concludes our discussion on non-locality for any $N$ party system. We now turn our attention to entanglement in $N$ qubit systems.
  %---------------------------------------------

\section{Entanglement}
\label{ent}
Ever since the concept of entanglement witness was introduced, a large number of them have been derived, initially for two qubit systems \cite{Guhne03} and later for $N$ qubit systems\cite{Guhne04, Toth05}. This is an ongoing project, and the derivations are based on diverse considerations. This section not only shows that  they  can all be arrived at through violation of classical probability rules, but also gives an explicit method of deriving many more such inequalities, without having to employ algebraic techniques. For three-qubit systems, we recover already known witnesses (inequalities 1 and 2 section (\ref{EntLin3})). To show the potential of the framework, we also  derive a new witness (inequality 3 in section (\ref{EntLin3})), with equal ease. In fact, in Appendix (\ref{Expansion}), we prove that any hermitian operator can be expanded in the overcomplete basis of pseudoprojection operators with nonnegative weights. Employing that expansion, the expectation value of any entanglement witness, which is necessarily a hermitian operator, can be written as a sum of  pseudoprobabilities. From this expansion, the corresponding violation of classical probability rule can be identified whenever the expectation value of entanglement witness becomes negative.

 We employ a common methodology for deriving entanglement inequalities which is outlined as follows.

\subsection{Methodology} 
\label{Methodology}
     While deriving conditions  for entanglement, we choose appropariate sums of pseudoprobabilties  and replace each of them by expectation of its parent pseudo-projection.  The choice of pseudoprobabilities has been made such that:
\begin{enumerate}
\item The sum of pseudoprobabilities involves two or more correlators between observables of the three or more qubits. The observables of each qubit do not commute, e.g., if the sum involves two terms $\langle a_1a_2a_3\rangle$ and $\langle a'_1a'_2a'_3\rangle$, where $a_i$ and $a'_i$ refer to observables corresponding to $i^{\rm th}$ qubit, then, $[a_i, a'_i]\neq 0$.
\item  The ensuing inequality is violated by all the fully separable states and is obeyed by at least one entangled state.
\end{enumerate}

The discussion to follow  is simplified by first representing the choice of observables geometrically. First of all, note that for a qubit, a dichotomic observable $a \equiv \vec{\sigma}\cdot\hat{a}$ is uniquely specified by the unit vector $\hat{a}$. 

  %---------------------------------------------

\subsubsection*{\bf Geometry}
\begin{figure}[htb!]
  \includegraphics[width=0.4\textwidth]{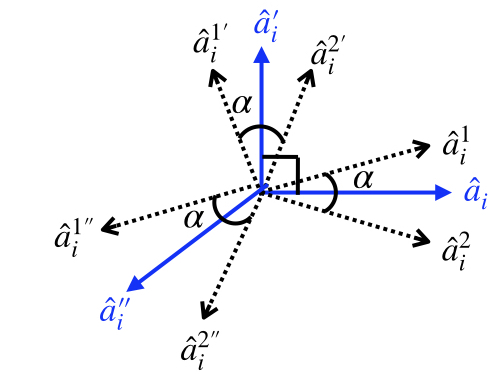}
  \caption{Directions chosen for pseudoprojections in construction of entanglement inequalities. $\hat{a}_i, \hat{a}'_i, \hat{a}''_i$: Orthonormal triplets in the space of $i^{\rm th}$ qubit that appear in the entanglement inequality (shown in blue). $\hat{a}^{1,2}_i, \hat{a}^{1',2'}_i, \hat{a}^{1'',2''}_i$: Directions used in the construction of pseudo-projection in the space of $i^{\rm th}$ qubit (shown in black).}
    \label{Geo}
\end{figure}

 All the cases discussed in this section have the same underlying geometry which we describe below:
\begin{enumerate}
\item  For each qubit, we have three sets  of doublets of observables. For the $i^{\rm th}$ qubit, we denote them by $\{ a^1_i, a^2_i\}$, $\{ a^{1'}_i, a^{2'}_i\}$, $\{ a^{1''}_i, a^{2''}_i\}$ (Recall $a^1_i\equiv \vec{\sigma}_i\cdot\hat{a}^1_i$). 
\item  The angles  between the observables in all the doublets take the same value, $\alpha$ -- which is the only free parameter, i.e., 
\begin{align}
{\rm Tr}(a_i^1a_i^2)\equiv &   {\rm Tr}(\vec{\sigma}_i\cdot\hat{a}^1_i\vec{\sigma}_i\cdot\hat{a}^2_i)=\hat{a}^1_i\cdot\hat{a}^2_i=\cos\alpha\nonumber\\
=&{\rm Tr}(\vec{\sigma}_i\cdot\hat{a}^{1'}_i\vec{\sigma}_i\cdot\hat{a}^{2'}_i)={\rm Tr}(\vec{\sigma}_i\cdot\hat{a}^{1''}_i\vec{\sigma}_i\cdot\hat{a}^{2''}_i)\nonumber
\end{align}. 

 \item  For each qubit, the normalised sums of vectors within each doublet form an orthonormal basis. For the $i^{\rm{th}}$ qubit, we represent the normalised sums of vectors within doublets $\{\hat{a}_1^1, \hat{a}_i^2\}, \{\hat{a}_1^{1'}, \hat{a}_i^{2'}\}$ and $\{\hat{a}_1^{1''}, \hat{a}_i^{2''}\}$ by $\hat{a}_i, \hat{a}'_i$ and $\hat{a}''_i$ respectively, i.e.,
 \begin{align}
 \hat{a}_i= \dfrac{\hat{a}_i^1+\hat{a}_i^2}{|\hat{a}_i^1+\hat{a}_i^2|} ;~\hat{a}'_i=\dfrac{ \hat{a}_i^{1'}+\hat{a}_i^{2'}}{|\hat{a}_i^{1'}+\hat{a}_i^{2'}|};~\hat{a}''_i= \dfrac{\hat{a}_i^{1''}+\hat{a}_i^{2''}}{|\hat{a}_i^{1''}+\hat{a}_i^{2''}|}.    
 \end{align}
\end{enumerate}
This  is completely depicted in figure (\ref{Geo}).

Unlike in  the case of nonlocality, we construct families of inequalities, some of  which also involve higher order moments of pseudo probabilities.
 Separable states can also display non-classical features, e.g., quantum discord. Thus, in order to construct sufficiency conditions for entanglement, we explicitly preclude separable states by fixing the range of $\alpha$ in each case.
  %---------------------------------------------

\subsection{Linear entanglement inequalities: $N=3$}
\label{EntLin3}
\subsubsection{ Inequality 1}
\label{Ineq1}
The first set of events which we consider leads to  entanglement inequalities involving only three body correlation terms. They involve the sum of pseudo probabilities,
\begin{align}
\label{Ent_continuous_1}
 {\cal P}^{1}_3 \equiv & \sum_{i=1}^2\big(\mathcal{P}({\cal E}_2^i; a'_3) + \mathcal{P}({\cal E}_2^{i+2}; {\bar a}'_3) +\mathcal{P}({\cal E}_2^{i+4}; a_3) + \mathcal{P}({\cal E}_2^{i+6}; {\bar a}_3)\big),
\end{align}
where, ${\cal E}^1_2 \equiv{\cal E}(a^1_1=a^2_1=a_2) ; {\cal E}^2_2 \equiv{\cal E}(a^{1'}_1=a^{2'}_1=a'_2)$. The other events are obtained via the transformations exhibited in Table (\ref{ent_3}).  

 \begin{table}
  \begin{center}
\begin{tabular}{ |M{2.5cm}|M{2.5cm}| } 
 \hline
 Set of events &  Observables  \\ 
 \hline
 ${\cal E}_2^{1,2} \rightarrow{\cal E}_2^{3,4}$ ${\cal E}_2^{5,6} \rightarrow{\cal E}_2^{7,8}$ & $a_2 (a'_2)\leftrightarrow {\bar a}_2 ({\bar a}'_2)$\\ 
 \hline
 ${\cal E}_2^{1,2} \rightarrow {\cal E}_2^{5,6}$  & $a_2\rightarrow  a'_2; ~a'_2\rightarrow {\bar a}_2$ ${\bar a}_2\rightarrow {\bar a}'_2; ~{\bar a}'_2\rightarrow a_2$\\
 \hline
\end{tabular}
\end{center}
  \caption{Transformation of ordered set of events and corresponding transformation of observables.\label{ent_3}}
 \end{table}
 Writing the pseudoprobabilities and imposing the nonclassicality condition ${\cal P}_3^1 <0,$ we arrive at the first family of entanglement  inequalities  

\begin{align}
\label{Ent_continuous}
E^{1}_3 \equiv \big\langle & 4 \cos\frac{\alpha}{2} +M_3  \big\rangle < 0;~0<\alpha\leq{\rm arccos}\Big(-\frac{7}{8}\Big),
\end{align}
where $M_3$ is the Mermin polynomial for the 3-qubit system. We now argue why the upper limit of $\alpha$ is ${\rm arccos}\Big(-\dfrac{7}{8}\Big)$. Note that $\langle M_3\rangle$ can take the minimum value $-1$ for separable states. To show this, without any loss of generality, we may choose:
\begin{align}
    a_1\equiv x_1;~a'_1\equiv y_1;~a_2\equiv x_2;~a'_2\equiv y_2;~a_3\equiv y_3;~a'_3\equiv x_3.\nonumber
\end{align}
With this choice, $M_3$ takes the following form:
\begin{align}
    M_3\equiv (x_1x_2+y_1y_2)x_3+(x_1y_2-y_1x_2)y_3,\nonumber
\end{align}
whose minimum expectation value for a fully separable three-qubit pure state $\dfrac{1}{8}(1+p_1)(1+p_2)(1+p_3)$ is $-1$. Since fully separable mixed states are convex sums of fully separable pure states, $\langle M_3\rangle$ has a  lower bound of $-1$ for fully separable mixed states as well. Thus, if we demand that $\big\langle 4\cos\dfrac{\alpha}{2}+M_3\big\rangle$ takes negative value only for entangled states, then $1 \leq 4\cos\dfrac{\alpha}{2} <4$. This, in turn, fixes the range of $\alpha$ to be  $0<\alpha\leq{\rm arccos}\Big(-\frac{7}{8}\Big)$. The bounds of $\alpha$ in the subsequent entanglement inequalities can be similarly found.

The detailed proof of inequality (\ref{Ent_continuous}) is given in Appendix (\ref{Derivation_E_3_1}).

  %---------------------------------------------

   \subsubsection{Inequality 2}
 We now refine the inequality $E^1_3$ (given in equation (\ref{Ent_continuous})) through the  inclusion of two-body correlation terms. Of interest is the sum, 
\begin{align}
 {\cal P}^{2}_3 \equiv & {\cal P}^1_3+\sum_{(ij)}^{} \mathcal{P}(a''_i  = a_{j}^{1''} =a_{j}^{2''}).
\end{align}
Here $(ij)$ represents cyclic permutation of $i,j = 1, 2, 3$.  Writing the pseudoprobabilities and imposing the nonclassicality condition ${\cal P}^{2}_3< 0$ implies the following inequality 
\begin{align}
\label{bisep}
 E^{2}_3  \equiv \big\langle & 7\cos\dfrac{\alpha}{2} + M_3+ a''_1a''_2 +a''_2a''_3 + a''_3a''_1\big\rangle < 0; ~0 < \alpha \leq  {\rm arccos}\Big(-\frac{31}{49}\Big).
\end{align}
  The range of $\alpha$ gets fixed by the demand that all the fully separable states violate the inequality $E_3^2$. The condition corresponding to $\alpha =  {\rm arccos}\big(-\frac{31}{49}\big)$ has been derived earlier as a $W$-- witness by  Ac\'{\i}n et al.\cite{Acin01} using algebraic approaches. 

%---------------------------------------------------------

  \subsubsection{Inequality 3}
  \label{New_ineq}
  It is possible to derive new inequalities from this framework. For example, we introduce the following combination which involves the same pseudoprobabilities as in ${\cal P}^{2}_3$ but with different weights:
  \begin{align}
     {\cal P}^{3}_3 \equiv & {\cal P}^1_3 + \dfrac{1}{3}\sum_{(ij)}^{} \mathcal{P}(a''_i  = a_{j}^{1''} =a_{j}^{2''}). 
  \end{align}
As before, writing the pseudoprobabilities and imposing the nonclassicality condition  ${\cal P}^{3}_3<0$ yields the following inequality for inseparability:
\begin{align}
\label{Complete_separability}
E^{3}_3 \equiv \Big\langle 5\cos\dfrac{\alpha}{2}+M_3
+&  \frac{1}{3} \big(a''_1a''_2 + a''_2a''_3 +a''_3a''_1\big)\Big\rangle < 0;~
0  < \alpha \leq {\rm arccos} \Big( -\frac{23}{25}\Big).
\end{align}
%In this case,  $0$. 
  The range of $\alpha$ gets fixed by the demand that all the completely separable states violate the inequality $E_3^3$.
 The detailed proof of inequality (\ref{Complete_separability}) is given in Appendix (\ref{Derivation_E_3_3}). Since the inequality is new, we also give the derivation for the range of $\alpha$ in the appendix.

The inequalities $E^1_3, E^2_3$ and $E^3_3$ can also be be derived using stabiliser formalism,  proposed in \cite{Toth05}. The present formalism successfully  traces back the underlying cause to classical probability rule violations.
  %---------------------------------------------

 \subsubsection{Inequality 4}
	Now we derive an entanglement inequality that involves correlation tensors  as well as local terms for those three qubit states which are  in the neighbourhood of the $W$-state. We start with the sum
 \begin{align}
 {\cal P}^{4}_3 \equiv\
  &\sum_{i=1}^3{\cal P}(a^{1''}_i, a^{2''}_i) \nonumber\\
  +&\sum_{i<j=1}^3\Big\{{\cal P}(a^{1''}_i= a^{2''}_i={\bar a}''_j)+2 \big\{{\cal P}(a^{1'}_i= a^{2'}_i=a'_j) + {\cal P}(a^{1}_i= a^{2}_i=a_j)\big\}\Big\}\nonumber\\
 +&3\big\{{\cal P}(a^{1''}_1=a^{2''}_1=a''_2;{\bar a}''_3)+{\cal P}(a^{1''}_1=a^{2''}_1={\bar a}''_2; a''_3)\big\}\nonumber\\
 +&\sum_{(ijk)}^{} \Big\{2\big\{\mathcal{P}(a^1_i = a^2_i =a_j; a''_k) + \mathcal{P}(a^{1'}_i = a^{2'}_i =a'_j; a''_k) \nonumber\\
 +& \mathcal{P}(a^1_i = a^2_i = {\bar a}_j; {\bar a}''_k) + \mathcal{P}(a^{1'}_i = a^{2'}_i = {\bar a}'_j; {\bar a}''_k)\big\}\Big\}.
 \end{align}
 Again, writing the pseudoprobabilities and imposing the nonclassicality condition the condition,  ${\cal P}^{4}_3 < 0$,  yields the following inequality for inseparability:
 \begin{align}
E^{4}_3 \equiv \Big\langle & 33\cos \dfrac{\alpha}{2} + \sum_{i=1}^3a''_i+\sum_{i<j=1}^3 \{2(a_ia_j+a'_ia'_j)-a''_ia''_j\}
-3a''_1a''_2a''_3+\nonumber\\
&2\sum_{(ijk)}^{}(a_ia_j+a'_ia'_j)a''_k\Big\rangle <0,\nonumber\\
&0 < \alpha \leq {\rm arccos}(-0.954) \approx \pi.
 \end{align}
 Here $(ijk)$ represents cyclic permutations of $i$, $j$ and $k$.  The range of $\alpha$ gets fixed by the demand that all the completely separable states violate the inequality $E_3^4$. The inequality corresponding to one particular value, $\alpha ={\rm arccos}(-0.689)$,  was  earlier derived in  \cite{Acin01} as a $W$-- witness.
  %---------------------------------------------

    \subsection{Bilinear entanglement inequalities: $N =3$}
\label{nonlinear}
        So far we have considered  violations of classical probability rules for linear combinations of pseudo-probabilities. We next turn our attention to  combinations involving blinear terms in pseudo-probabilities.
        
        %---------------------------------------------------------
        
    \subsubsection{Inequality 1}
 Consider, first, the sum of products of pseudoprobabilities:
\begin{align}
  {\cal S}^{1}_3 \equiv
 & \Big\{\mathcal{P}({\cal E}_2^1; a_3) + \mathcal{P}({\cal E}_2^2 ; a_3) + \mathcal{P}({\cal E}_2^3; {\bar a}_3) + \mathcal{P}({\cal E}_2^4; {\bar a}_3)\Big\} \nonumber\\
 \times & \Big\{\mathcal{P}({\cal E}_2^1; {\bar a}_3) + \mathcal{P}({\cal E}_2^2 ; {\bar a}_3) + \mathcal{P}({\cal E}_2^3; a_3) + \mathcal{P}({\cal E}_2^4; a_3)\Big\} \nonumber\\
 + &\big\{\mathcal{P}({\cal E}_2^5; a'_3) + \mathcal{P}({\cal E}_2^6; {\bar a}'_3)+ \mathcal{P}({\cal E}_2^7; {\bar a}'_3) + \mathcal{P}({\cal E}_2^8; a'_3) \big\}\nonumber\\
 \times & \big\{\mathcal{P}({\cal E}_2^5; {\bar a}'_3)  + \mathcal{P}({\cal E}_2^6; a'_3)+\mathcal{P}({\cal E}_2^7; a'_3)+ \mathcal{P}({\cal E}_2^8; {\bar a}'_3) \big\},
\end{align}
where, the events are given  by, 
\begin{align}
\label{E1-8}
& {\cal E}_2^1 \equiv {\cal E}(a^1_1 = a^2_1 =a_2);~{\cal E}_2^2 \equiv {\cal E}(a^{1'}_1 = a^{2'}_1 =a'_2);\nonumber\\
& {\cal E}_2^3 \equiv {\cal E}(a^1_1 = a^2_1 = {\bar a}_2);~{\cal E}_2^4 \equiv {\cal E}(a^{1'}_1 = a^{2'}_1 ={\bar a}'_2)\nonumber\\
& {\cal E}_2^5 \equiv {\cal E}(a^1_1 = a^2_1 =a'_2);~{\cal E}_2^6 \equiv {\cal E}(a^{1'}_1 = a^{2'}_1 =a_2);\nonumber\\
& {\cal E}_2^7 \equiv {\cal E}(a^1_1 = a^2_1 = {\bar a}'_2);~{\cal E}_2^8 \equiv {\cal E}(a^{1'}_1 = a^{2'}_1 ={\bar a}_2).
\end{align}
 Writing the pseudoprobabilities and imposing the nonclassicality condition
 ${\cal S}^{1}_3< 0$ yields the family of inequalities,
\begin{align}
\mathbb{E}^{1}_3\equiv  8\cos^2\dfrac{\alpha}{2} &- \big\langle (a_1a_2 +a'_1a'_2)a_3\big\rangle^2- \big\langle (a'_1a_2-a_1a'_2)a'_3\big\rangle^2 <0\nonumber\\
 &0 < \alpha \leq {\rm arccos} \Big( -\frac{3}{4}\Big)
\end{align}
  The range of $\alpha$ gets fixed by the demand that all the separable states violate the inequality $\mathbb{E}_3^1$.
 %Furthermore,  $$. 
   %---------------------------------------------

 \subsubsection{Inequality 2}
The second bilinear combination that we consider includes contribution from two-body correlation terms as well:
\begin{align}
\label{Ent_continuous_11}
 {\cal S}^{2}_3 \equiv 
  & \big\{\mathcal{P}({\cal E}_2^1; a_3) + \mathcal{P}({\cal E}_2^3; {\bar a}_3) \big\}\big\{\mathcal{P}({\cal E}_2^1; {\bar a}_3) + \mathcal{P}({\cal E}^3; a_3)\big\} \nonumber\\
 +& \big\{\mathcal{P}({\cal E}_2^2; a_3) +\mathcal{P}({\cal E}_2^4; {\bar a}_3) \big\}\big\{ \mathcal{P}({\cal E}_2^2; {\bar a}_3)  + \mathcal{P}({\cal E}^4; a_3) \big\}\nonumber\\
  +&\big\{\mathcal{P}({\cal E}_2^5; a'_3) + \mathcal{P}({\cal E}_2^7; {\bar a}'_3)\big\}\big\{\mathcal{P}({\cal E}_2^5; {\bar a}'_3)  + \mathcal{P}({\cal E}_2^7; a'_3) \big\}\nonumber\\
  +& \big\{\mathcal{P}({\cal E}_2^6; {\bar a}'_3) + \mathcal{P}({\cal E}_2^6; a'_3)\big\}\big\{\mathcal{P}({\cal E}_2^8; a'_3)  +  \mathcal{P}({\cal E}_2^8; {\bar a}'_3)\big\}\nonumber\\
 +& \sum_{(ij)}^{}\mathcal{P}(a''_i  = a_j^{1''} =a_j^{2''})\mathcal{P}({\bar a}''_i  = a_j^{1''} =a_j^{2''}),
\end{align}
where $\{{\cal E}_2^1,\cdots,{\cal E}_2^8\}$ have been defined in equation (\ref{E1-8}). $(ij)$ represents cyclic permutations of $1, 2,3$. Writing the pseudoprobabilities and imposing the  nonclassicality condition ${\cal S}^2_3< 0$ implies,
\begin{align}
\mathbb{E}^{2}_3& \equiv  7 \cos^2 \dfrac{\alpha}{2} - \langle a_1''a_2''\rangle^2  - \langle a''_2a''_3 \rangle^2 -\langle a''_3a''_1 \rangle^2 \nonumber\\
-& \langle a_1a_2a_3\rangle^2 -\langle a_1a'_2a'_3\rangle^2- \langle a'_1a_2a'_3\rangle^2 - \langle a'_1a'_2a_3 \rangle^2 <0,\nonumber\\
&0 < \alpha \leq {\rm arccos} \Big( -\frac{1}{7}\Big).
\end{align}
%with the constraint  $$.
   The range of $\alpha$ gets fixed by the demand that all the fully separable states violate the inequality $\mathbb{E}_3^2$.
The inequality  corresponding to the particular value $\alpha = {\rm arccos} \big( -\frac{1}{7}\big)$ has been derived in \cite{Guhne04} by invoking  bounds on variances of observables for  separable states vis-a vis fully tripartite entangled states. 
   
  %---------------------------------------------

    \subsection{Linear entanglement inequalities for $N$-- qubit system}
    \label{Ent_linN}
     We now generalise our results by constructing  an entanglement inequality involving $N$--body correlations. The correlation terms in the inequality are just the highest rank tensors that would occur in  the  $N$--qubit GHZ state. We start with  the sum of pseudoprobabilities,
         \begin{align}
            {\cal P}_N &= \sum_{i=1}^{\lambda}\big\{{\cal P}({\cal E}_{N-1}^i; a'_N)+{\cal P}({\cal E}_{N-1}^{i+\lambda}; {\bar a}'_N)
            +{\cal P}({\cal E}_{N-1}^{i+2\lambda}; a_N)+{\cal P}({\cal E}_{N-1}^{i+3\lambda}; {\bar a}_N)\big\};~N \geq 3.
         \end{align}
where   $\lambda = 2^{2N-5}$ and the events ${\cal E}_{N-1}^i; i \in \{1, \cdots , \lambda\}$ are to be extracted, recursively,  from those pseudoprobabilities that  underlie  the  entanglement inequality for  $(N-1)$ qubits. The basic pseudo probabilities for  two-qubits are given by
\begin{align}
 {\cal P}_2 &= {\cal P}(a^1_1=a^2_1=a_2)
 +{\cal P}(a^{1'}_1=a^{2'}_1=a'_2).
\end{align}
Thus, ${\cal E}_2^ 1 \equiv{\cal E}(a^1_1=a^2_1=a_2) $ and ${\cal E}_2^2\equiv{\cal E}(a^{1'}_1=a^{2'}_1=a'_2)$.
 \begin{table}[ht]
  \begin{center}
\begin{tabular}{ |M{3.5cm}|M{5cm}| } 
 \hline
Ordered set of events & Observables \\ 
 \hline
$\{{\cal E}_{N-1}^1, \cdots , {\cal E}_{N-1}^{\lambda}\} \rightarrow$ $\{{\cal E}_{N-1}^{\lambda+1}, \cdots , {\cal E}_{N-1}^{2\lambda}\}$ 
$\{{\cal E}_{N-1}^{2\lambda+1}, \cdots , {\cal E}_{N-1}^{3\lambda}\} \rightarrow$
$\{{\cal E}_{N-1}^{3\lambda+1}, \cdots ,{\cal E}_{N-1}^{4\lambda}\}$ & $a_{N-1}\leftrightarrow {\bar a}_{N-1}$; $a'_{N-1}\leftrightarrow {\bar a}'_{N-1} $\\
 \hline
 $\{{\cal E}_{N-1}^1, \cdots , {\cal E}_{N-1}^{\lambda}\} \rightarrow$ & $a_{N-1}\rightarrow { a}'_{N-1}$; {$a'_{N-1}\rightarrow {\bar a}_{N-1} $}\\ 
 {$\{{\cal E}_{N-1}^{2\lambda+1}, \cdots , {\cal E}_{N-1}^{3\lambda}\}$} & ${\bar a}_{N-1}\rightarrow {\bar a}'_{N-1}; ~{\bar a}'_{N-1}\rightarrow { a}_{N-1}$\\ 
 \hline
\end{tabular}
\end{center}
  \caption{Transformation of ordered set of events and corresponding transformation of observables.\label{ent_N}}
 \end{table}

     Writing all the pseudoprobabilites and imposing the nonclassicality condition ${\cal P}_N<0$, the  entanglement inequality that follows has the form 
     \begin{align}
 E_N  \equiv \Big\langle 2^{N-1}\cos\dfrac{\alpha}{2}+M_N\Big\rangle <0, 
     \end{align}
    Imposing the condition $E_N\geq 0$ for all the completely separable states fixes  $0< \alpha\leq {\rm arccos}\Big(-\frac{2^{2N-3}-1}{2^{2N-3}}\Big)$, which approaches the value $ \pi$ as $N \rightarrow \infty$.  The range of $\alpha$ gets fixed by the demand that  all the separable states violate the inequality $E_N$.
     
     If $\alpha$ is left unrestricted, all the states with nonzero correlation tensor, i.e., having terms like $x_1x_2x_3, x_1y_2y_3, y_1x_2y_3$ and $y_1y_2x_3$ in the density matrix,  will be detected to be nonclassical.
\subsubsection{Comments on generalisation}
 It is by now clear that, by following the method that we have employed, numerous entanglement inequalities for a multi-party system can be constructed, by including  correlations in the subsystems. The strength of the framework lies in the fact that no extra concept, other than violation of a classical probability rule, is required. 
 The task gets further facilitated by a result which we prove in the Appendix (\ref{Expansion}): that any observable admits an expansion, with non-negative coefficients,  in the overcomplete basis provided by a set of elementary pseudo projections.
 Specialisation to entanglement is accomplished by carving out suitable regions in the parameter space.

  %---------------------------------------------
  
\section{Examples}
\label{Illustration}
\subsection{Three-qubit GHZ state with white noise}
We now present the results for the noise resistance of the state, $|GHZ_3\rangle=  \frac{1}{\sqrt{2}}(|000\rangle + |111\rangle$, with respect to the inequalities derived above. The state is given by 
\begin{equation}
\rho_p =  p |GHZ_3\rangle \langle GHZ_3| + (1-p) \frac{1}{8}I
\end{equation}
where $I$ is the identity operator.   The parameter $p$ determines the purity of the state. The ranges of $p$ for which $\rho_p$ is detected to be nonlocal or entangled by different nonlocality and entanglement inequalities are shown in tables (\ref{Illustration_t1}) and (\ref{Illustration_t}) respectively.  
  \begin{table}[ht]
  \begin{center}
\begin{tabular}{ |c|c|c| } 
 \hline
 Inequality &  Range of $p$ \\ 
 \hline
 ${\cal I}_{S_3}$  &  $\frac{1}{\sqrt{2}}< p \leq 1$\\
 \hline
 ${\cal I}_{M_3}$ & $\frac{1}{2}<  p \leq 1$\\ 
 \hline
 \end{tabular}
\end{center}
  \caption{Range of $p$ for which the state $\rho_p$ is detected to be nonlocal by Svetlichny and Mermin inequalities for hybrid local-nonlocal and completely factorisable local hidden variable models.\label{Illustration_t1}}
  \end{table}
 \begin{table}[ht]
  \begin{center}
\begin{tabular}{ |c|c|c| } 
 \hline
 Inequality &  Range of $p$ \\ 
 \hline
 $E^1_3$ & $\frac{1}{4}< p \leq 1$\\
 \hline
 $E^2_3$ & $\frac{3}{5}< p \leq 1$\\
 \hline
 $E^3_3$ & $\frac{1}{5}< p \leq 1$\\ 
  \hline
  $\mathbb{E}_3^1$ & $\frac{1}{2\sqrt{2}}< p \leq 1$\\
  \hline
  $\mathbb{E}_3^2$ & $\sqrt{\frac{3}{7}}< p \leq  1$\\
  \hline
\end{tabular}
\end{center}
  \caption{Range of $p$ for which $\rho$ is detected to be entangled by different entanglement inequalities.\label{Illustration_t}}
 \end{table}
  For graphical representation, we show the corresponding ranges of $p$ in figures (\ref{Nonlocality_ran}) and (\ref{Entanglement_ran}) respectively. Evidently, the inequality $E^3_3$ detects the entangled states in the range $\dfrac{1}{5} <p\leq 1$. 
 The entry in the fifth row, corresponding to  $E^3_3$,  has been obtained in  \cite{Guhne10} by imposing conditions on the entries of density matrix  of completely separable and entangled states using concavity arguments.
%-----------------------------
\begin{figure}
 \begin{tikzpicture}
 \draw [black,thick,-](0,0) -- (2.8284,0);
 \draw [black,thick, -](2.8284,0) -- (4,0);
 \draw [black,thick,-](2,0) -- (2.8284,0);
\draw [black,thick, ->](0.1,-0.2) -- (0.6,-0.2);
\draw [brown,thick,-](2,0.15) -- (2,-0.15);
\draw [green,thick,-](2.82,0.15) -- (2.82,-0.15);
\node (a1) at (0,-0.9) [label={0}]{};
\node (a1) at (4,-0.9) [label={1}]{};
\node (a1) at (0,-0.5) [label={$p$}]{};
\node (a1) at (2,0) [label={$Q$}]{};
\node (a1) at (2.8284,-0.9) [label=$\frac{1}{\sqrt{2}}$]{};
\node (a1) at (2,-0.9) [label=$\frac{1}{2}$]{};
\node (a1) at (2.8284,0.05) [label={$P$}]{};
\node (a1) at (0,0.05) [label=$O$]{};
\node (a1) at (4,0.05) [label=$A$]{};
 \end{tikzpicture}
 \caption{$AP$ and $AQ$: Ranges of $p$ for which the state $\rho$ is detected to be nonlocal by three-party Svetlichny and Mermin inequalities (${\cal I}_{S_3}$ and ${\cal I}_{M_3}$) respectively for hybrid local-nonlocal and completely factorisable local hidden variable models.}
  \label{Nonlocality_ran}
 \end{figure}
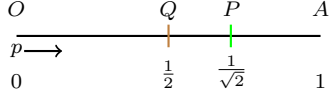
 \begin{figure}
 \begin{tikzpicture}
 \draw [black,thick,-](0,-0.5) -- (6.7075,-0.5);
 \draw [black,thick, -](6.7075,-0.5) -- (10,-0.5);
  \draw [green,thick, -](6.7075,-0.3) -- (6.7075,-0.7);
  \draw [black,thick,-](6,-0.5) -- (6.7075,-0.5);
  \draw [yellow,thick,-](6,-0.3) -- (6,-0.7);
  \draw [black,thick, -](3.53,-0.5) -- (6,-0.5);
  \draw [red,thick, -](3.53,-0.3) -- (3.53,-0.7);
  \draw [black,thick,-](2.5,-0.5) -- (3.53,-0.5);
  \draw [black,thick,-](0,-0.5) -- (2,-0.5);
  \draw[violet,thick,-](2.5,-0.3) -- (2.5,-0.7);
    \draw[blue,thick,-](2,-0.3) -- (2,-0.7);
    \node (a1) at (10,-1.5) [label=1]{};
    \node (a1) at (0,-1.5) [label=0]{};
\node (a1) at (0,-1.1) [label={$p$}]{};
\draw[black,thick,->](0.25,-0.8) -- (1.0,-0.8);
\node (a1) at (6.5465,-1.55) [label=$\sqrt{\frac{3}{7}}$]{};
\node (a1) at (6.7,-0.45) [label=$L_5$]{};
\node (a1) at (6,-1.5) [label=$\frac{3}{5}$]{};
\node (a1) at (6.1,-0.45) [label=$L_4$]{};
\node (a1) at (3.5355,-1.55) [label=$\frac{1}{2\sqrt{2}}$]{};
\node (a1) at (3.65,-0.45) [label=$L_3$]{};
\node (a1) at (2.5,-1.55) [label=$\frac{1}{4}$]{};
\node (a1) at (2.6,-0.45) [label=$L_2$]{};
\node (a1) at (2,-1.55) [label=$\frac{1}{5}$]{};
\node (a1) at (2.1,-0.45) [label=$L_1$]{};
\node (a1) at (0,-0.45) [label=$O$]{};
\node (a1) at (10,-0.45) [label=$A$]{};
 \end{tikzpicture}
\caption{$AL_1$, $AL_2$, $AL_3$, $AL_4$,  $AL_5$: Ranges of $p$ for which the state $\rho$ is detected to be entangled by entanglement inequalities $E_3^3, E_3^1, \mathbb{E}^1_3, {E}^2_3, \mathbb{E}^2_3$ respectively.}
\label{Entanglement_ran}
 \end{figure}
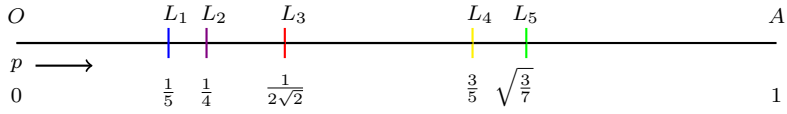
 \subsection{$N$-qubit GHZ state with white noise}
 The state $\rho_p = p |GHZ_N\rangle \langle GHZ_N| + \frac{1-p}{2^N}I$ is  detected to be entangled by the inequality $E_N$ in the range $ \frac{1}{2^{N-1}}< p \leq 1$.  
  %---------------------------------------------

\section{Conclusion}
\label{Conclusion}
In conclusion, we have proved that  violation of any nonlocality inequality is equivalent to a sum of pseudoprobabilities assuming a negative value and any hermitian operator can be written as a sum of pseudoprojections with non-negative weights.  The difference between both the features gets reflected in that the  pseudoprobabilities in the former do not have their origin in any particular theory. Entanglement, being a nonclassical feature of quantum mechanics, the underlying pseudoprobabilities are expectations of pseudoprojections.  Using this method, we have recovered  a multitude of well known results for non locality and entanglement in multi party/multi qubit systems derived earlier by using several different means. Furthermore, we have derived a new family of entanglement inequalities in section (\ref{New_ineq}). Finally, we have  indicated how many more nonclassicality conditions may be derived without a need to introducing new concepts or techniques.

  %---------------------------------------------

%\section*{Acknowledgement}
\begin{acknowledgements}
It is a pleasure to thank Rajni Bala for fruitful discussions.  We thank the anonymous referees whose comments have helped to enhance the quality of manuscript to a large extent. Sooryansh  thanks the Council for Scientific and Industrial Research (Grant no. -09/086 (1278)/2017-EMR-I) for funding his research.
\end{acknowledgements}

%---------------------------------------------------------
\section*{Author Contribution Statement}
All the authors contributed equally in all respects.
 \section*{Conflict of interest}
 The authors declare that they have no conflict of interest.
\appendix
\section*{Appendix}
\section{Violation of a nonlinear nonlocality inequality implies violation of a classical probability rule}
\label{NonlinearNonlocProof}
In this appendix, we prove that violation of any nonlinear nonlocality inequality is equivalent to the nonexistence of an underlying nonnegative  pseudoprobability scheme. We show it for a bipartite system, but the proof admits a straightforward generalisation to multipartite systems as well.\\

\noindent{\it Condition for locality:}
Let $\{C_{\alpha}\}$ and $\{D_{\beta}\}$ be the sets of $N_1$ and $N_2$ observables for the first and the second subsystems of a bipartite system respectively. The respective sets of outcomes of the observables are $\{c^{i_{\alpha}}_{\alpha}\}$ and $\{d^{j_{\beta}}_{\beta}\}$. The most general nonlinear nonlocality inequality obeyed by all the local hidden variable models reads as,
\begin{align}
\label{NonlinNL}
{\cal I}_{\rm NL}\equiv    \sum_{abc\cdots} w_{abc\cdots}({\cal P}_{a})^{m_{a}}({\cal P}_b)^{m_{b}}({\cal P}_c)^{m_{c}}\cdots \leq R,
\end{align}
where all ${\cal P}_{a}, {\cal P}_{b}$ and ${\cal P}_{c}$ are of the form ${\cal P}(C_{\alpha}=c^{i_{\alpha}}_{\alpha}, D_{\beta}=d^{j_{\beta}}_{\beta})$, for some $(C_{\alpha}, D_{\beta}, c_{\alpha}^{i_{\alpha}}, d_{\beta}^{j_{\beta}})$ and $m_{a}, m_{b}, m_{c}, \cdots \geq 1$. 
The inequality has been so written that all $w_{abc\cdots} \geq 0$. All the inequalities can be brought in this form. For example, if there is a term ${\cal P}^l~ (l > 1),$ with a negative coefficient, then it can be rewritten as $1-{\cal P}'(1+{\cal P}+{\cal P}^2+\cdots + {\cal P}^{l-1})$, where ${\cal P}'$ is the probability of the complementary event.\\

\noindent {\it Lemma:} Let ${\cal M}_1, {\cal M}_2,\cdots$ be the sets of all the mutually consistent events $\{C_{\alpha}=c_{\alpha}^{i_{\alpha}}, D_{\beta}=d_{\beta}^{j_{\beta}}\}$. For a given ${\cal M}_k$, let ${\cal N}_k$ represent the set of the coefficients $w_{abc\cdots}$ of those terms in inequality (\ref{NonlinNL}) that contain the probabilities for the mutually consistent events belonging to ${\cal M}_k$. If the sum of all $w_{abc\cdots} \in {\cal N}_k$ is represented by $R_k$, then,
\begin{align}
\label{Lemmaequiv}
    R \geq \underset{\{{\cal N}_k\}}{\rm max}~ R_k,
\end{align}
where the maximum is taken over all ${\cal N}_1, {\cal N}_2, \cdots$.\\
\noindent{\it Proof:} For a given $k$, there always exists a local hidden variable model in which all the mutually consistent events belonging to the set ${\cal M}_k$ can be assigned unit probability (and all the other events inconsistent with the events belonging to ${\cal M}_k$ have zero probability). Since $k$ is an arbitrary label, equation (\ref{Lemmaequiv}) holds.\\

\noindent{\it Theorem 1A:} Violation of inequality (\ref{NonlinNL}) implies violation of a classical probability rule.\\
{\it Proof:}  The proof follows from similar arguments as in section (\ref{Proof_NL}). Consider the expression,
\begin{align}
    R-{\cal I}_{\rm NL} =R- \sum_{abc\cdots} w_{abc\cdots}({\cal P}_{a})^{m_{a}}({\cal P}_{b})^{m_b}({\cal P}_c)^{m_c}\cdots.
    \label{NLexpression}
\end{align} 
Since all the joint probabilities  in the second term of equation (\ref{NLexpression}) are of the form ${\cal P}(C_{\alpha}=c_{\alpha}^{i_{\alpha}}, D_{\beta}=d_{\beta}^{j_{\beta}})$, we rewrite each of them as $\sideset{}{'}\sum {\cal P}(\{C_{\alpha}=c_{\alpha}^{i_{\alpha}}\}, \{D_{\beta}=d_{\beta}^{j_{\beta}}\})$, where $\sideset{}{'}\sum$ represents summation over all the outcomes $\{c_{\alpha}^{i_{\alpha}}\}$ and $\{d_{\beta}^{j_{\beta}}\}$ except $c_{\alpha}^{i_{\alpha}}$ and $d_{\beta}^{j_{\beta}}$. Let ${\rm max}(m_{a}+m_{b}+m_{c}+\cdots) = M$, where the maximum is taken over all $a, b, c, \cdots$. Then, we insert the following identity in the first term of equation (\ref{NLexpression}):
\begin{align}
    \Big\{\sum_{\substack{i_1\cdots i_{N_1}\\j_1\cdots j_{N_2}}}{\cal P}(\{C_{\alpha}=c_{\alpha}^{i_{\alpha}}\},  \{D_{\beta}=d_{\beta}^{j_{\beta}}\})\Big\}^M =1.
\end{align}
It is straightforward to see that the coefficients of all the joint probabilities of all orders in eqaution (\ref{NLexpression}), after these substitutions, are nonnegative. If, in addition, the joint probabilities are also nonnegative, it follows that,
\begin{align}
    R-{\cal I}_{\rm NL} \geq 0 \implies {\cal I}_{\rm NL} \leq R,
\end{align}
which agrees with the locality condition given in equation (\ref{NonlinNL}). Thus, in order that the locality condition gets violated, some of the joint probabilities have to turn negative. This implies that in place of a joint probabilities, we  have pseudoprobabilities that may assume negative values as well.

%--------------------
\section{Derivation of the three-party Svetlichny inequality}
\label{Svet_Calc}
In this section, we detail  the derivation of the three-party Svetlichny inequality. First, we write the joint probabilities of the events in terms of expectation values as follows:
\begin{align}
{\cal E}_2^{\alpha} &\equiv{\cal E}({\bar A}_1=A'_1=A_2)\vee{\cal E}(A_1=A'_1=A'_2)\nonumber\\
&\equiv{\cal E}({\bar A}_1A'_1;A_2)\vee{\cal E}(A_1{\bar A}'_1;{\bar A}_2)\vee{\cal E}(A_1A'_1;A'_2)\vee{\cal E}({\bar A}_1{\bar A}'_1;{\bar A}'_2)\nonumber\\
 {\cal P}({\cal E}^{\alpha}_2;A'_3)&=\dfrac{1}{16}\Big\langle\Big\{(1-A_1A'_1+(-A_1+A'_1)A_2-A_1+A'_1+A_2)\Big\}\nonumber\\
&+ (1-A_1A'_1+(-A_1+A'_1)A_2+A_1-A'_1-A_2)\nonumber\\
&+ (1+A_1A'_1+(A_1+A'_1)A'_2+A_1+A'_1+A'_2)\nonumber\\
&+ (1+A_1A'_1+(A_1+A'_1)A'_2-A_1-A'_1-A'_2)\Big\}(1+A'_3)\Big\rangle\nonumber\\
&=\dfrac{1}{8}\Big\langle\Big\{2+(-A_1+A'_1)A_2+(A_1+A'_1)A'_2\Big\}(1+A'_3)\Big\rangle\nonumber\end{align}
Thus,
\begin{align}
\label{Event1}
    {\cal P}({\cal E}^{\alpha}_2; A'_3)+{\cal P}({\cal E}^{\alpha'}_2; {\bar A}'_3) &=\dfrac{1}{4}\Big\langle\Big\{2+\{(-A_1+A'_1)A_2+(A_1+A'_1)A'_2\}A'_3\Big\}\Big\rangle
\end{align}
Similarly,
\begin{align}
\label{Event2}
     {\cal P}({\cal E}^{\beta}_2; A_3)+{\cal P}({\cal E}^{\beta'}_2; {\bar A}_3) &=\dfrac{1}{4}\Big\langle\Big\{2+\{(-A'_1+A_1)A'_2+(A_1+A'_1)A_2\}A_3\Big\}\Big\rangle
\end{align}
Adding the joint probabilities in equations (\ref{Event1}) and (\ref{Event2}), 
\begin{align}
\label{Event12}
{\cal P}^S_3\equiv&{\cal P}({\cal E}^{\alpha}_2; A'_3)+{\cal P}({\cal E}^{\alpha'}_2; {\bar A}'_3)+
    {\cal P}({\cal E}^{\beta}_2; A_3)+{\cal P}({\cal E}^{\beta'}_2; {\bar A}_3)\nonumber\\
    =\dfrac{1}{4}\Big\langle&\big\{4+\{(-A_1+A'_1)A_2+(A_1+A'_1)A'_2\}A'_3+\{(-A'_1+A_1)A'_2+(A_1+A'_1)A_2\}A_3\big\}\Big\rangle
\end{align}
Imposing the demand ${\cal P}^S_3<0$, the following inequality emerges
\begin{align}
\label{Condition1}
    \langle S_3\rangle <-4.
\end{align}
Interchanging ${\bar A}_1$ and ${\bar A}'_1$ with ${A}_1$ and ${A}'_1$ respectively in equation (\ref{Event12}) and imposing the nonclassicality condition, we obtain the complementary condition $\langle S_3\rangle>4$, which, together with the condition given in equation (\ref{Condition1}) yields $|\langle S_3\rangle|>4$.
%-------------------------------------------------
\section{Expansion of any hermitian operator as sum of pseudoprojections}
\label{Expansion}
In this appendix, we show how an {\it arbitrary} hermitian operator $\hat{O}$ in dimension $d$ can be expanded as sum of pseudoprojections with non-negative weights.  Let the hermitian operator $\hat{O}$ be written in the basis spanned by generalised Pauli matrices as:
\begin{align}
\hat{O} =& w + x^{ij}X_{ij} + y^{ij}Y_{ij} +z^{l}Z_{l}\nonumber\\
=&w+ (x_+^{ij}+x_-^{ij})X_{ij} + (y_+^{ij}+y_-^{ij})Y_{ij} +(z_+^{l}+z_-^l)Z_{l}\nonumber\\
=&w+ (x_+^{ij}+x^{ij}_-)X_{ij} + (y^{ij}_++y^{ij}_-)Y_{ij} +(z^{i,l+1}_++z^{i,l+1}_-)Z_{i,l+1}
\end{align}
where, 
\begin{align}
    & f^{ij}_{\pm}= f^{ij} \Theta(\pm f^{ij}).
\end{align}
 $\Theta (\cdot)$ represents Heaviside unit step function. Repeated indices are all summed over and,
\begin{align}
X_{ij}^{kl} =& \delta_{ki}\delta_{lj} + \delta_{li}\delta_{kj}~{\rm for}~1\leq i<j\leq d,\nonumber\\
Y_{ij}^{kl} =& -i (\delta_{ki}\delta_{lj} - \delta_{li}\delta_{kj})~{\rm for}~1\leq i<j\leq d,\nonumber\\
Z_l =&  \sqrt{\frac{2}{l(l+1)}}\Big(\sum_{j=1}^l|j\rangle\langle j|-l |l+1\rangle\langle l+1|\Big);~1 \leq~l~\leq~d-1,\nonumber\\
=&  \sqrt{\frac{2}{l(l+1)}} \Big\{ (|1\rangle\langle 1|-|l+1\rangle\langle l+1|) +\cdots + (|l\rangle\langle l|-|l+1\rangle\langle l+1|)\Big\}\nonumber\\
=& \sqrt{\frac{2}{l(l+1)}} \Big\{ Z_{1,l+1} +\cdots + Z_{l,l+1}\Big\}.
\end{align}
where 
\begin{align}
    Z_{i,l+1}=|i\rangle\langle i|-|l+1\rangle\langle l+1
    |;~i=1,\cdots,l\nonumber
\end{align}
We introduce two doublets of unit vectors ($\hat{p}_1$, $\hat{p}_2$) and ($\hat{q}_1$, $\hat{q}_2$) such that the included angle between the two vectors of each doublet is $\theta$. Let $(\hat{p}_1+\hat{p}_2), (\hat{p}_1-\hat{p}_2)$ and $(\hat{q}_1+\hat{q}_2)$ form an orthonormal triad, then we can always choose $(\hat{p}_1+\hat{p}_2)\parallel\hat{x}$, $(\hat{p}_1-\hat{p}_2)\parallel\hat{y}$, $(\hat{q}_1+\hat{q}_2)\parallel\hat{z}$. Following are the short-hand notations for pseudoprojections for different joint events:
\begin{align}
& {\bf \Pi}_{p^{ij}_1p^{ij}_2} \equiv{\bf \Pi}^{X_{ij}}_{+};~~~~ {\bf \Pi}_{{\bar p}^{ij}_1{\bar p}^{ij}_2}\equiv{\bf \Pi}^{X_{ij}}_{-}\nonumber\\
& {\bf\Pi}_{{p}^{ij}_1{\bar p}^{ij}_2}\equiv{\bf \Pi}^{Y_{ij}}_{+};~~~~ {\bf \Pi}_{{\bar p}^{ij}_1p^{ij}_2}\equiv{\bf \Pi}^{Y_{ij}}_{-}\nonumber\\
& {\bf \Pi}_{q^{ij}_1q^{ij}_2} \equiv{\bf \Pi}^{Z_{ij}}_{+};~~~~ {\bf \Pi}_{{\bar q}^{ij}_1{\bar q}^{ij}_2}\equiv{\bf \Pi}^{Z_{ij}}_-, 
\end{align}
Here $p_1^{ij} \equiv \vec{\sigma}^{ij}\cdot\hat{p}_1$ and so on.  

One more pseudoprojection ${\bf \Pi}_{r^{ij}_1r^{ij}_2r^{ij}_3}$, representing the joint event $\vec{\sigma}_{ij}\cdot\hat{r}_1=+1;~\vec{\sigma}_{ij}\cdot\hat{r}_2=+1;~\vec{\sigma}_{ij}\cdot\hat{r}_3=+1$, will also be required; where $\hat{r}_i$ are coplanar and at an included angle of $2\pi/3$ with each other,
\begin{align}
 {\bf \Pi}_{r^{ij}_1r^{ij}_2r^{ij}_3}&=\dfrac{1}{3!}\big(\pi_{r^{ij}_1}\pi_{r^{ij}_2}\pi_{r^{ij}_3}+\pi_{r^{ij}_1}\pi_{r^{ij}_3}\pi_{r^{ij}_2}+\pi_{r^{ij}_2}\pi_{r^{ij}_1}\pi_{r^{ij}_3}+\pi_{r^{ij}_2}\pi_{r^{ij}_3}\pi_{r^{ij}_1}\nonumber\\
 &+\pi_{r^{ij}_3}\pi_{r^{ij}_1}\pi_{r^{ij}_2}+\pi_{r^{ij}_3}\pi_{r^{ij}_2}\pi_{r^{ij}_1}\big) =-1_{ij}/16,
 \nonumber
\end{align}
  where $1_{ij}^{kl}=\delta_{ik}\delta_{il}+\delta_{jk}\delta_{jl}$.\\
The expressions of different pseudoprojections are as follows:
\begin{align}
& {\bf \Pi}_{\pm}^{X_{ij}} = \dfrac{1}{2}\cos\dfrac{\theta}{2}\Big(\cos\dfrac{\theta}{2}1_{ij} \pm X_{ij}\Big);
~~ {\bf \Pi}_{\pm}^{Y_{ij}} = \dfrac{1}{2}\sin\dfrac{\theta}{2}\Big(\sin\dfrac{\theta}{2}1_{ij} \pm Y_{ij}\Big)\nonumber\\
& {\bf \Pi}^{Z_{i,l+1}}_{\pm} = \dfrac{1}{2}\cos\dfrac{\theta}{2}\Big(\cos\dfrac{\theta}{2}1_{i,l+1} \pm Z_{i,l+1}\Big).
\end{align} 
The operator $\hat{O}$ can be expanded in terms of pseudoprojections and the expansion coefficients are as follows:
%%-----------------------------------------------------------------------
\begin{align}
\hat{O} =& c+
2\sec\frac{\theta}{2}\Big(x^{ij}_+{\bf \Pi}^{X_{ij}}_{+} + |x^{ij}_-|{\bf \Pi}^{X_{ij}}_{-}
+z^{'i,l+1}_{+}{\bf \Pi}^{Z_{i,l+1}}_+ + |z^{'i,l+1}_{-}|{\bf \Pi}^{Z_{i,l+1}}_-\Big)\nonumber\\
+& 2{\rm cosec}\frac{\theta}{2}\Big(y^{ij}_{+}{\bf \Pi}^{Y_{ij}}_{+} + |y^{ij}_{-}|{\bf \Pi}^{Y_{ij}}_{-}\Big) 
\end{align}
where
\begin{align}
z^{'}_+ &= \sqrt{\frac{2}{l(l+1)}}z_+; z^{'}_- = \sqrt{\frac{2}{l(l+1)}}z_{-},\nonumber\\
c =& \Big\{w
-(x_{+}^{ij}+|x_{-}^{ij}|)\cos\frac{\theta}{2}1_{ij}-(y_{+}^{ij}+|y_{-}^{ij}|)\sin\frac{\theta}{2}1_{ij}-
(z^{'i,l+1}_++|z^{'i,l+1}_-|)\cos\frac{\theta}{2}1_{i,l+1}\Big\}\nonumber\\
=&\Big\{\frac{w}{d-1}\sum_{i<j=1}^d\Big(\sec^2\frac{\theta}{2}({\bf\Pi}^{X_{ij}}_++{\bf \Pi}^{X_{ij}}_-)\Theta(w)
+ 16{\bf\Pi}_{r^{ ij}_1r^{ij}_2r^{ij}_3}\Theta(-w)\Big)+\nonumber\\
+&16\Big((x_+^{ij}+|x^{ij}_-|)\cos\frac{\theta}{2}+(y_+^{ij}+|y^{ij}_-|)\sin\frac{\theta}{2}\Big){\bf\Pi}_{r^{ij}_1r^{ij}_2r^{ij}_3}\nonumber\\
+&16(z^{'i,l+1}_++|z^{'i,l+1}_-|)\cos\frac{\theta}{2}{\bf\Pi}_{r_1^{i,l+1}r_2^{i,l+1}r_3^{i,l+1}}\Big\}
\end{align}
Here, $\Theta(w)$ represents Heaviside step function.
%---------------------------------------------------
\section{Derivation of the entanglement inequality $E_3^1$}
\label{Derivation_E_3_1}
The sum of pseudoprobabilities underlying this inequality is given in equation (\ref{Ent_continuous_1}), which is as follows:
\begin{align}
{\cal P}^{1}_3 \equiv & \sum_{i=1}^2\big(\mathcal{P}({\cal E}_2^i; a'_3) + \mathcal{P}({\cal E}_2^{i+2}; {\bar a}'_3) +\mathcal{P}({\cal E}_2^{i+4}; a_3) + \mathcal{P}({\cal E}_2^{i+6}; {\bar a}_3)\big),
\end{align}
where, ${\cal E}^1_2 \equiv{\cal E}(a^1_1=a^2_1=a_2) ; {\cal E}^2_2 \equiv{\cal E}(a^{1'}_1=a^{2'}_1=a'_2)$.
The psudoprobabilities corresponding to different events in equation (\ref{Ent_continuous_1}) are as follows
\begin{align}
    {\cal P}({\cal E}^1_2) = {\cal P}(a^1_1=a^2_1=a_2) &= \Big\langle\dfrac{1}{4}\cos\dfrac{\alpha}{2}\Big(\cos\dfrac{\alpha}{2}+a_1\Big) (1+a_2)+\dfrac{1}{4}\cos\dfrac{\alpha}{2}\Big(\cos\dfrac{\alpha}{2}-a_1\Big) (1-a_2)\Big\rangle\nonumber\\\nonumber\\
    &=\Big\langle \dfrac{1}{4}\cos\dfrac{\alpha}{2}\Big(2\cos\dfrac{\alpha}{2}+2a_1a_2\Big)\Big\rangle
\end{align}
Similarly,
\begin{align}
 {\cal P}({\cal E}_2^2)&=\Big\langle \dfrac{1}{4}\cos\dfrac{\alpha}{2}\Big(2\cos\dfrac{\alpha}{2}+2a'_1a'_2\Big)\Big\rangle\nonumber\\
    {\cal P}({\cal E}_2^3)&=\Big\langle \dfrac{1}{4}\cos\dfrac{\alpha}{2}\Big(2\cos\dfrac{\alpha}{2}-2a_1a_2\Big)\Big\rangle;~~~{\cal P}({\cal E}_2^4)&=\Big\langle \dfrac{1}{4}\cos\dfrac{\alpha}{2}\Big(2\cos\dfrac{\alpha}{2}-2a'_1a'_2\Big)\Big\rangle\nonumber\\
    {\cal P}({\cal E}_2^5)&=\Big\langle \dfrac{1}{4}\cos\dfrac{\alpha}{2}\Big(2\cos\dfrac{\alpha}{2}+2a_1a'_2\Big)\Big\rangle;~~~{\cal P}({\cal E}_2^6)&=\Big\langle \dfrac{1}{4}\cos\dfrac{\alpha}{2}\Big(2\cos\dfrac{\alpha}{2}-2a'_1a_2\Big)\Big\rangle\nonumber\\  
     {\cal P}({\cal E}_2^7)&=\Big\langle \dfrac{1}{4}\cos\dfrac{\alpha}{2}\Big(2\cos\dfrac{\alpha}{2}-2a_1a'_2\Big)\Big\rangle;~~~{\cal P}({\cal E}_2^8)&=\Big\langle \dfrac{1}{4}\cos\dfrac{\alpha}{2}\Big(2\cos\dfrac{\alpha}{2}+2a'_1a_2\Big)\Big\rangle
\end{align}
Substituting these explicit forms of the pseudoprobabilities in eqaution (\ref{Ent_continuous}), we obtain
\begin{align}
    &{\cal P}({\cal E}_2^1;a_3')+ {\cal P}({\cal E}_2^2;a_3')+ {\cal P}({\cal E}_2^3;{\bar a}_3')+ {\cal P}({\cal E}_2^4;{\bar a}_3')+\nonumber\\
    &{\cal P}({\cal E}_2^5;a_3)+ {\cal P}({\cal E}_2^6;a_3)+ {\cal P}({\cal E}_2^7;{\bar a}_3)+ {\cal P}({\cal E}_2^8;{\bar a}_3)\nonumber\\
    =&\dfrac{1}{4}\cos\dfrac{\alpha}{2}\Big\langle\Big[\Big\{\Big(\cos\dfrac{\alpha}{2}+a_1a_2\Big)+\Big(\cos\dfrac{\alpha}{2}+a'_1a'_2\Big)\Big\}(1+a'_3)\nonumber\\
    +&\Big\{\Big(\cos\dfrac{\alpha}{2}-a_1a_2\Big)+\Big(\cos\dfrac{\alpha}{2}-a'_1a'_2\Big)\Big\}(1-a'_3)\nonumber\\
    +&\Big\{\Big(\cos\dfrac{\alpha}{2}+a_1a'_2\Big)+\Big(\cos\dfrac{\alpha}{2}-a'_1a_2\Big)\Big\}(1+a_3)\nonumber\\
    +&\Big\{\Big(\cos\dfrac{\alpha}{2}-a_1a'_2\Big)+\Big(\cos\dfrac{\alpha}{2}+a'_1a'_2\Big)\Big\}(1-a_3)\Big\}\Big]\Big\rangle\nonumber\\
    =&\dfrac{1}{4}\cos\dfrac{\alpha}{2}\Big\langle 8\cos\dfrac{\alpha}{2}+2(a_1a_2+a'_1a'_2)a'_3+2(a_1a'_2-a'_1a_2)a_3\Big\rangle
    \label{final}
\end{align}
Imposing the nonclassicality condition on equation (\ref{final}), i.e., demanding the sum of pseudoprobabilities to be negative, the inequality $E_3^1$ emerges.

\section{Proof of the entanglement inequality $E_3^3$}
\label{Derivation_E_3_3}
We have the following relation:
\begin{align}
    {\cal P}(a_i{''}=a_j^{1''}=a_j^{2''})=& {\cal P}(a_i{''}=a_j^{1''}=a_j^{2''}=+1)+ {\cal P}(a_i{''}=a_j^{1''}=a_j^{2''}=-1)\nonumber\\
    =&\dfrac{1}{4}\cos\dfrac{\alpha}{2}\Big\langle\Big(\cos\dfrac{\alpha}{2}+a_i''\Big)\Big(1+a_j''\Big)+\Big(\cos\dfrac{\alpha}{2}-a_i''\Big)\Big(1-a_j''\Big)\Big\rangle\nonumber\\
    =&\dfrac{1}{4}\cos\dfrac{\alpha}{2}\Big(2\cos\dfrac{\alpha}{2}+2a''_ia''_j\Big)\Big\rangle=\dfrac{1}{2}\cos\dfrac{\alpha}{2}\Big(\cos\dfrac{\alpha}{2}+a''_ia''_j\Big)\Big\rangle
    \label{Second Terms}
\end{align}
Thus, plugging the values from equations (\ref{final}) and (\ref{Second Terms}), 
\begin{align}
     &{\cal P}({\cal E}_2^1;a_3')+ {\cal P}({\cal E}_2^2;a_3')+ {\cal P}({\cal E}_2^3;{\bar a}_3')+ {\cal P}({\cal E}_2^4;{\bar a}_3')\nonumber\\
    +&{\cal P}({\cal E}_2^5;a_3)+ {\cal P}({\cal E}_2^6;a_3)+ {\cal P}({\cal E}_2^7;{\bar a}_3)+ {\cal P}({\cal E}_2^8;{\bar a}_3)+\frac{1}{3}\sum_{(ij)}{\cal P}(a''_i=a^{1''}_j=a^{2''}_j)\nonumber\\
    =&\dfrac{1}{2}\cos\dfrac{\alpha}{2}\Big\langle 5\cos\dfrac{\alpha}{2} + M_3+\dfrac{1}{3}(a''_1a''_2+a''_2a''_3+a''_3a''_1)\Big\rangle
    \label{final1}
\end{align}
Imposing the nonclassicality condition on equation (\ref{final1}), i.e., demanding the sum of pseudoprobabilities to be negative, the inequality $E_3^3$ emerges, i.e.,
\begin{align}
    \Big\langle 5\cos\dfrac{\alpha}{2}+M_3+\dfrac{1}{3}(a''_1a''_2+a''_2a''_3+a''_3a''_1)\Big\rangle <0\nonumber
\end{align}
In order to fix the range of $\alpha$, note that, without any loss of generality, we may choose,
\begin{align}
    a_1\equiv x_1;~a'_1\equiv y_1;~a''_1\equiv z_1;~a_2\equiv x_2;~a'_2\equiv y_2;~a''_2\equiv z_2;~a_3\equiv y_3;~a'_3\equiv x_3; a''_3\equiv z_3. \nonumber
\end{align}
With this choice,
\begin{align}
    M_3+\dfrac{1}{3}(a''_1a''_2+a''_2a''_3+a''_3a''_1) &\equiv (x_1x_2+y_1y_2)x_3+(x_1y_2-y_1x_2)y_3\nonumber\\
    &+\dfrac{1}{3}(z_1z_2+z_2z_3+z_3z_1),\nonumber
\end{align}
whose expectation value for a pure separable three -qubit state $\frac{1}{8}(1+p_1)(1+p_2)(1+p_3)$ ($|p_1|=|p_2|=|p_3|=1$) is given by,
\begin{align}
  &(p_{1x}p_{2x}+p_{1y}p_{2y})p_{3x} +(p_{1x}p_{2y}-p_{1y}p_{2x})p_{3y}+\dfrac{1}{3}(p_{1z}p_{2z}+p_{2z}p_{3z}+p_{3z}p_{1z}).  
\end{align}
If $\vec{p}_i \equiv (p_{ix}, p_{iy}, p_{iz}) \equiv (\sin\theta_i\cos\phi_i, \sin\theta_i\sin\phi_i, \cos\theta_i); i \in \{1, 2, 3\}$, then,
\begin{align}
  &(p_{1x}p_{2x}+p_{1y}p_{2y})p_{3x} +(p_{1x}p_{2y}-p_{1y}p_{2x})p_{3y}+\dfrac{1}{3}(p_{1z}p_{2z}+p_{2z}p_{3z}+p_{3z}p_{1z})\nonumber\\ 
  =&\sin\theta_1\sin\theta_2\sin\theta_3\cos(\phi_1-\phi_2+\phi_3)+\dfrac{1}{3}(\cos\theta_1\cos\theta_2+\cos\theta_2\cos\theta_3+\cos\theta_3\cos\theta_1)\nonumber\\
   \geq &-\sin\theta_1\sin\theta_2\sin\theta_3+\dfrac{1}{3}(\cos\theta_1\cos\theta_2+\cos\theta_2\cos\theta_3+\cos\theta_3\cos\theta_1)\nonumber\\
   =&\dfrac{1}{3}\big(-3\sin\theta_1\sin\theta_2\sin\theta_3+\cos\theta_1\cos\theta_2+\cos\theta_2\cos\theta_3+\cos\theta_3\cos\theta_1\big)\nonumber\\
   \geq& \dfrac{1}{3}\big(\cos (\theta_1+\theta_2) + \cos (\theta_2+\theta_3) + \cos (\theta_3 + \theta_1)\big) \geq -1.
\end{align}
Thus, in order that the inequality $\Big\langle 5\cos\alpha+M_3+\dfrac{1}{3}(a''_1a''_2+a''_2a''_3+a''_3a''_1)\Big\rangle <0$ gets violated by all separable states, $ 1 \leq 5\cos\dfrac{\alpha}{2} < 5 $, which, in turn fixes the range of $\alpha$ to be $0 < \alpha \leq {\rm arccos} \Big(-\dfrac{23}{25}\Big)$.

%------------------------------

%-----------------------------

%-------------------------------

% Authors must disclose all relationships or interests that 
% could have direct or potential influence or impart bias on 
% the work: 
%

% BibTeX users please use one of
%\bibliographystyle{spbasic}      % basic style, author-year citations
%\bibliographystyle{spmpsci}      % mathematics and physical sciences
\bibliographystyle{spphys}       % APS-like style for physics
%\bibliography{bibliography}

% name your BibTeX data base

% Non-BibTeX users please use
%\begin{thebibliography}{}
%
% and use \bibitem to create references. Consult the Instructions
% for authors for reference list style.
%
%\bibitem{RefJ}
% Format for Journal Reference
%Author, Article title, Journal, Volume, page numbers (year)
% Format for books
%\bibitem{RefB}
%Author, Book title, page numbers. Publisher, place (year)
% etc
%\end{thebibliography}

\end{document}